\documentclass[%
superscriptaddress,
preprintnumbers,
amsmath,amssymb,
aps,
prfluids,
longbibliography,
notitlepage,
floatfix
]{revtex4-2}

\usepackage{graphicx}

\usepackage[labelformat=simple]{subcaption}
\usepackage{bm} 

\usepackage{siunitx} 

\usepackage[colorlinks=true]{hyperref}

\newcommand{\rd}{\mathrm{d}}
\newcommand{\ri}{\mathrm{i}}
\newcommand{\re}{\mathrm{e}}
\DeclareMathOperator{\Real}{Re}

\allowdisplaybreaks 

\begin{document}

\title{Oscillatory flows in compliant conduits at arbitrary Womersley number}

\author{Shrihari D.\ Pande}
\affiliation{School of Mechanical Engineering, Purdue University, West Lafayette, Indiana 47907, USA}

\author{Xiaojia Wang}
\altaffiliation[Present address: ]{Department of Mathematics, University of Michigan, Ann Arbor, Michigan 48109, USA.}
\affiliation{School of Mechanical Engineering, Purdue University, West Lafayette, Indiana 47907, USA}

\author{Ivan C.\ Christov}
\thanks{Corresponding author.}
\email{christov@purdue.edu}\homepage{http://tmnt-lab.org}
\affiliation{School of Mechanical Engineering, Purdue University, West Lafayette, Indiana 47907, USA}
\affiliation{Department of Computer Science, University of Nicosia, 46 Makedonitissas Avenue, CY-2417, Nicosia, Cyprus}

\date{\today}

\begin{abstract}
We develop a theory of fluid--structure interaction (FSI) between an oscillatory Newtonian fluid flow and a compliant conduit. We consider the canonical geometries of a 2D channel with a deformable top wall and an axisymmetric deformable tube. Focusing on the hydrodynamics, we employ a linear relationship between wall displacement and hydrodynamic pressure, which has been shown to be suitable for a leading-order-in-slenderness theory. The slenderness assumption also allows the use of lubrication theory, and the flow rate is related to the pressure gradient (and the tube/wall deformation) via the classical solutions for oscillatory flow in a channel and in a tube (attributed to Womersley). Then, by two-way coupling the oscillatory flow and the wall deformation via the continuity equation, a one-dimensional nonlinear partial differential equation (PDE) governing the instantaneous pressure distribution along the conduit is obtained, without \textit{a priori} assumptions on the magnitude of the oscillation frequency (\textit{i.e.}, at arbitrary Womersley number). We find that the cycle-averaged pressure (for harmonic pressure-controlled conditions) deviates from the expected steady pressure distribution, suggesting the presence of a streaming flow. An analytical perturbative solution for a weakly deformable conduit is obtained to rationalize how FSI induces such streaming. In the case of a compliant tube, the results obtained from the proposed reduced-order PDE and its perturbative solutions are validated against three-dimensional, two-way-coupled direct numerical simulations. We find good agreement between theory and simulations for a range of dimensionless parameters characterizing the oscillatory flow and the FSI, demonstrating the validity of the proposed theory of oscillatory flows in compliant conduits at arbitrary Womersley number.
\end{abstract}

\maketitle

\section{Introduction}
\label{sec:intro}

A decade ago, \citet{VOB10} noted that ``[t]he field of pulsatile microfluidics is largely unexplored.'' Since then, numerous applications of pulsatile flows to microfluidics have been realized \cite{DDS20}, including for lab-on-a-chip technologies \cite{Huhetal10,BI14}. However, despite the fact that deformations of the flow conduit due to pulsatile flow are observed \cite{BULHLB09,RWJ21}, and used for flow control and shaping \cite{LESKUBL09,WKB10,CRFZULB13,BPCOJ22} and biomimicry \cite{RDC19,PBG22}, amongst other applications \cite{XWZZW21}, a theory of the fluid--structure interaction (FSI) between an oscillatory internal flow and the walls of a compliant conduit is lacking for the lubrication (low Reynolds number) limit relevant to microfluidics. Generally, these types of FSI problems fall within the subject of \emph{elastohydrodynamics} \cite{G01}. Previous work on FSI of oscillatory internal flows in compliant conduits has primarily focused on the inertial (moderate Reynolds number) limit relevant to flow in large blood vessels \cite{P80,GJ04}. Similarly, previous work on elastohydrodynamic has focused on applications relevant to tribology, in which the pressures generated are so extreme that even conventionally ``hard'' materials are deformed by the fluid flow \cite{G01,LME11}.

In the biofluid mechanics context, the linearized elastohydrodynamic problem has been extensively analyzed (discussed in textbooks \cite{A16,SS12,Z00}), dating back to Womersley's classical work \cite{W55,W55a}. Here, by ``linearized elastohydrodynamic problem'' we mean that the vessel wall deformations are considered infinitesimal and they do not appreciably change the cross-sectional area of the conduit itself (see also the discussion in \cite{FAVF20}). (It should be noted that \citet{W55} realizes this limitation on p.~212 and attempts to find a correction due to the nonlinear flow--deformation coupling in \cite[\S5]{W55} and \cite[Sec.~VIII]{W57c}. However, this point does not seem to have been fully fleshed out since; interestingly, ``part II'' of \cite{W55} does not appear to have been published.) This assumption is also made in more recent microfluidics works \cite{UBH11}. From these linearized FSI problems' solutions, concepts such as \emph{resistance}, \emph{capacitance} and \emph{inductance} can be defined for hydraulic circuits, in analogy to electrical circuits (see, \textit{e.g.}, \cite{B08,K10}). In cardiovascular fluid mechanics, these models are known as \emph{windkessel} models \cite{F97} (and include several generalizations as well as limitations \cite{WLW04}). Indeed, microfluidic experiments highlight limitations of simplified linear theories \cite{MF04}, and require the fitting of the resistance, inductance, and capacitance values \cite{PJBSKDL16} (or introducing ``nonconventional'' damping \cite{XHSA20}) to predict the resonant peaks in microfluidic circuits.

One reason for these discrepancies between Womersley's classical approach (and the subsequent lumped-parameter windkessel models) and experiments (or direct numerical simulations) is that the pressure (or pressure gradient) is generally considered to be known. Meanwhile, for flow in long compliant conduits, the pressure gradient is unknown \textit{a priori}; it must be found self-consistently from the coupled solution of the elastohydrodynamic problem (typically as a solution to a nonlinear ordinary differential equation) \cite{RK72,C21}. In this work, we present a theory of the nonlinear pressure--flow relationship for pulsatile flow in a compliant conduit without \textit{a priori} assumptions on the magnitude of the oscillation frequency.  We term this lack of assumption on the frequency as ``arbitrary Womersley number'' in Table~\ref{tb:literature}, to distinguish from the limit of steady flow (zero oscillation frequency) or the case of ``small'' oscillation frequency (suitably defined through the Womersley number).

To this end, we consider two canonical geometries of interest: a two-dimensional (2D) channel (Sec.~\ref{sec:channel}), and a three-dimensional (3D) but axisymmetric tube (Sec.~\ref{sec:tube}). The lubrication approximation is reviewed for each of these (Secs.~\ref{sec:lubrication_channel} and \ref{sec:lubrication_tube}). In the presence of two-way coupled FSI, the fluid's momentum equation cannot be solved exactly; nevertheless, we motivate a von K\'{a}rm\'{a}n--Pohlhausen-type closure using the known axial velocity profiles (and volumetric flow rate) for pulsatile flow (Secs.~\ref{sec:flow_solution_channel} and \ref{sec:flow_solution_tube}). The flow rate expression, and suitable models for the flow-induced deformation of the geometries for the pressure's evolution (Secs.~\ref{sec:deformation_model_channel} and \ref{sec:deformation_model_tube}), are coupled via the conservation of mass equation to obtain (nonlinear) partial differential equations (PDEs) for the pressure's evolution (Secs.~\ref{sec:reduced_model_channel} and \ref{sec:reduced_model_tube}). Restricting to \emph{oscillatory} flows, which do not have a mean-flow component (unlike the more general pulsatile flows), the governing PDEs for the pressure can be solved analytically for weakly deformable conduits (Secs.~\ref{sec:weak_FSI_channel} and \ref{sec:weak_FSI_tube}). More generally, we solve the PDEs numerically (Secs.~\ref{sec:results_chan} and \ref{sec:Results_Tube}). Comparing the analytical and numerical results, we demonstrate a \emph{streaming} phenomenon due to FSI in incompressible oscillatory flow in a compliant conduit. The streaming effect is observed in both geometries considered, and certain universal features are identified via the perturbative analysis (Secs.~\ref{sec:weak_FSI_channel} and \ref{sec:weak_FSI_tube}). For the case of an axisymmetric tube, we also perform 3D direct numerical simulations of the two-way coupled FSI to validate our reduced-order model (Sec.~\ref{sec:Results_Tube}). Conclusions and avenues for future work are discussed in Sec.~\ref{sec:conclusion}. 

\begin{table}
    \centering
    \resizebox{\textwidth}{!}{
    \begin{tabular}{l@{\quad} l@{\quad} l@{\quad} l@{\quad} l}
    \hline\hline
    Reference & Focus & Geometry & Womersley number & Compliance number \\
    \hline
    \citet{W55,W57c} & Theory & Tube & Arbitrary & $=0$, wall inertia considered\\   
    \citet{CLMT05} & Theory & Tube & $\mathcal{O}(1)$ & $\ll1$ \\
    \citet{CTGMHR06} & Theory \& experiment, viscoelastic tube & Tube & $\mathcal{O}(1)$ & $\ll1$ \\
    \citet{WKB10} & Theory \& experiment & 3D channel & $\ll 1$ & $=0$, lumped-parameter model,\\ 
    &&&& deformation seen in experiment\\
    \citet{VOB10} & Theory \& experiment & Tube & Arbitrary & $=0$, lumped-parameter model\\
    \citet{SS12} & Theory & Tube & Arbitrary & Compliance idealized as slip\\
    \citet{EG14} & Theory & Tube & $\ll 1$ & $=0$ \\
    \citet{BBG17} & Theory, non-Newtonian fluid & Tube & $\ll 1$ & $=0$ \\
    \citet{RDC19} & Experiment, non-Newtonian fluid & Tube & $0$ to $2.17$ & Unknown, deformation observed\\
    \citet{TG19} & Theory & 2D channel & Arbitrary & $=0$, wall inertia considered\\
    \citet{AC20} & Theory, compressible fluid & Tube & $\ll 1$ & $\ll 1$ \\
    & \& viscoelastic tube &&& \\
    \citet{VJ20} & Experiment & 3D channel & $1.5$ to $15$ & No deformation observed\\
    This work & Theory \& 3D direct simulation & 2D channel \& Tube & Arbitrary & Not assumed small \\
    \hline\hline
    \end{tabular}
    }
    \caption{A chronological selection of studies on low-Reynolds-number, oscillatory flows in long, slender compliant conduits. Unless otherwise noted, studies involve Newtonian fluids and linearly elastic walls. Wall inertia is neglected unless otherwise stated. The Womersley number, defined in Tables~\ref{table:param_2D} and \ref{table:param_AS} below, quantifies the order of magnitude of unsteady inertial forces compared to viscous forces. The compliance number, also defined in Tables~\ref{table:param_2D} and \ref{table:param_AS} below, quantifies the order of magnitude of the hydrodynamic pressure to the wall's elastic resistance. The papers classified as ``theory'' above often also necessitate the numerical solution of a reduced model (as in the present work).}
    \label{tb:literature}
\end{table}

\section{Oscillatory flow in a two-dimensional channel with a compliant wall}
\label{sec:channel}

\subsection{Governing equations: scaling and lubrication approximation}
\label{sec:lubrication_channel}

\begin{figure}
    \centering
    \includegraphics[width=0.6\textwidth]{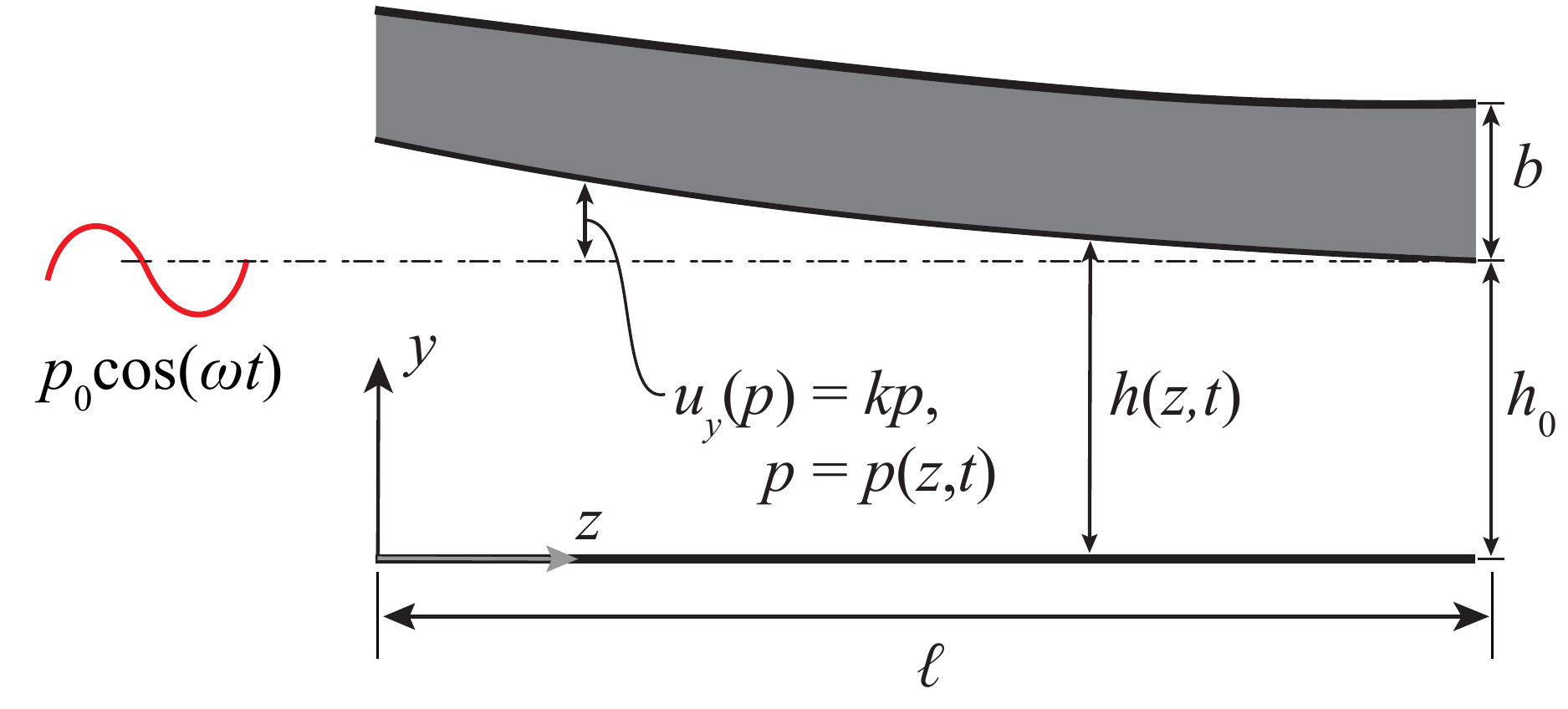}
    \caption{Schematic of a 2D compliant microchannel, indicating key quantities and notation for the geometry.}
    \label{fig:schematic_channel}
\end{figure}

Consider a two-dimensional (2D) channel in the $(y,z)$ plane as shown in Fig.~\ref{fig:schematic_channel}. The width $w$ in the spanwise $x$-direction (into the page, not shown) is so large that the flow may be considered 2D and independent of $x$. Assume a Newtonian fluid with its dynamic viscosity and density being $\mu_f$ and $\rho_f$, respectively. The flow is driven by an oscillatory pressure (frequency $\omega$, amplitude $p_0$) imposed at the inlet, and the channel is open to the atmosphere at its outlet. Neglecting any body forces, the mass and momentum conservation equations for this fluid flow \cite{panton} are
\begin{subequations}
\begin{align}
    \underbrace{\frac{\partial v_y}{\partial y}}_{\mathcal{O}(1)}+ \underbrace{\frac{\partial v_z}{\partial z}}_{\mathcal{O}(1)}&=0,\label{eq:com_2D}\\
    \underbrace{\rho_f\frac{\partial v_y}{\partial t}}_{\mathcal{O}(\epsilon^2 {\alpha}^2)}
    + \underbrace{\rho_fv_y\frac{\partial v_y}{\partial y}}_{\mathcal{O}(\epsilon^3 Re)}+\underbrace{\rho_fv_z\frac{\partial v_y}{\partial z}}_{\mathcal{O}(\epsilon^3 Re)}&=\underbrace{\mu_f\frac{\partial^2 v_y}{\partial y^2}}_{\mathcal{O}(\epsilon^2)}+\underbrace{\mu_f\frac{\partial^2 v_y}{\partial z^2}}_{\mathcal{O}(\epsilon^4)}-\underbrace{\frac{\partial p}{\partial y}}_{\mathcal{O}(1)},\label{eq:y_momentum_2D}\\
    \underbrace{\rho_f\frac{\partial v_z}{\partial t}}_{\mathcal{O}({\alpha}^2)}
    + \underbrace{\rho_fv_y\frac{\partial v_z}{\partial y}}_{\mathcal{O}(\epsilon Re)}+\underbrace{\rho_fv_z\frac{\partial v_z}{\partial z}}_{\mathcal{O}(\epsilon Re)}&=\underbrace{\mu_f\frac{\partial^2 v_z}{\partial y^2}}_{\mathcal{O}(1)}+\underbrace{\mu_f\frac{\partial^2 v_z}{\partial z^2}}_{\mathcal{O}(\epsilon^2)}-\underbrace{\frac{\partial p}{\partial z}}_{\mathcal{O}(1)}.\label{eq:z_momentum_2D}
\end{align}\label{eq:iNS_2D}\end{subequations}
The scales used to determine the orders of magnitude of terms in Eqs.~\eqref{eq:iNS_2D} are given in Table~\ref{table:scales_2D}. Three dimensionless numbers ($\epsilon$, $Re$ and $\alpha$) arise. They are defined in Table~\ref{table:param_2D}, where their typical values are also given.

\begin{table}[hb]
\begin{tabular}{l@{\qquad} l@{\qquad} l@{\qquad} l@{\qquad} l@{\qquad} l@{\qquad} l}
    \hline\hline
    Variable & $t$ & $y$ & $z$ & $v_y$ & $v_z$ & $p$\\
    \hline
    Scale & ${2 \pi}/{\omega}$ & $h_0$ & $\ell$ & $\epsilon \mathcal{V}_z$ & $\mathcal{V}_z = {\epsilon h_0 p_0}/{\mu_f}$ & $p_0$\\
    \hline\hline
    \end{tabular}
\caption{The scales for the variables in the 2D incompressible Navier--Stokes equations~\eqref{eq:iNS_2D}. Note that we have used the lubrication-theory scales for  $v_y$, $v_z$ based on the pressure scale imposed by $p_0$ \cite{S17_LH}.}
\label{table:scales_2D}
\end{table}

As usual, the Reynolds number $Re$ gauges the order of magnitude of \emph{convective} inertial forces compared to viscous forces. (As discussed in \cite{S17_LH}, the \emph{effective} Reynolds number $\epsilon Re$ is actually the relevant quantity herein, on which we make assumptions in Sec.~\ref{sec:flow_solution_channel} below.) Meanwhile, the Womersley number $\alpha$ measures the order of magnitude of \emph{unsteady} inertial forces compared to viscous forces. Although we consider the case of $h = h(z,t)$, we assume that $\max_z h(z,t)$ is always on the order of $h_0$, to be consistent with the lubrication approximation, which we now make.

\subsection{Flow solution at arbitrary Womersley number}
\label{sec:flow_solution_channel}
For a microfluidic system of interest, as discussed in Sec.~\ref{sec:intro}, Table~\ref{table:param_2D} presents the typical values of the relevant dimensional and dimensionless parameters. Thus, we are led to consider the limit of $\epsilon\ll 1$, $\epsilon Re\ll 1$, and ${\epsilon^2{\alpha}^2}\ll1$. This limit is the well-known \emph{lubrication approximation} \cite{panton,S17_LH}, but observe that ${\epsilon^2{\alpha}^2}\ll1$ allows the Womersley number $\alpha$ to be $\mathcal{O}(1)$, or even larger, within the same approximation. 
In this regime, Eq.~\eqref{eq:z_momentum_2D} reduces to
\begin{equation}
    \rho_f\frac{\partial v_z}{\partial t}=\mu_f\frac{\partial^2 v_z}{\partial y^2}-\frac{\partial p}{\partial z},
    \label{eq:z_lubrication_2D}
\end{equation}
subject to no slip along the channel walls, $v_z(y=0,t)=v_z(y=h,t)=0$. Observe that Eq.~\eqref{eq:z_lubrication_2D}, arising from the lubrication approximation for a deformable channel of variable height $h$, is the same as the axial momentum equation in a rigid channel of constant height $h_0$, reduced identically for 2D unidirectional flow \cite{panton}. Indeed, the channel height $h$ does not have to be constant under the lubrication approximation, as long as it \emph{varies slowly} \cite{VD87}.

\begin{table}
    \centering
    \begin{tabular}{l@{\quad} l@{\quad} l@{\quad} l}
    \hline\hline
    Quantity & Notation & Typical value & Units \\
    \hline
    Channel's length & $\ell$ & $1.0$ & \si{\centi\meter} \\
    Channel's undeformed height & $h_{0}$ & $10$ to $100$ & \si{\micro\meter} \\
    Top wall's thickness & $b$ & $10$ to $100$ & \si{\micro\meter} \\
    Solid's Young's modulus & $E$ & $0.1$ to $1$ & \si{\mega\pascal}  \\
    Solid's Poisson's ratio & $\nu_s$ & $0.49$ to $0.5$ & -- \\
    Solid's density & $\rho_s$ & $1.0 \times 10^3$ & \si{\kilo\gram\per\meter\tothe{3}} \\
    Fluid's density & $\rho_f$ &$1.0\times10^{3}$ & \si{\kilo\gram\per\meter\tothe{3}} \\
    Fluid's dynamic viscosity & $\mu$ & $1.0\times10^{-3}$ & \si{\pascal\second} \\
    Pressure pulse amplitude & $p_0$ & $1$ to $10$ & \si{\kilo\pascal} \\
    Pressure pulse frequency & $\omega/2 \pi$ & $1$ to $10^3$ & Hz \\
    \hline
    Channel's height-to-length aspect ratio & $\epsilon={h_0}/{\ell}$ & $0.001$ to $0.01$ &  --\\
    Reynolds number & $Re={\rho_f \epsilon h_0^2 p_0}/{\mu_f^2}$  & $0.1$ to $100$ &  --\\
    Womersley number & ${\alpha}=h_0\sqrt{\rho_f \omega/\mu_f}$  & $0.03$ to $7$ &  --\\
    FSI (or, compliance) number & $\beta=k p_0/h_0 $ &  $0.001$ to $0.1$ & -- \\
    Solid's Strouhal number & $St_s ={\rho_s b \mathcal{U}\omega^2}/(4\pi^2 p_0) $ & 
    $\approx 10^{-9}$ to $10^{-6}$ & -- \\
    Fluid's Strouhal number & $St_f = \ell \mu_f \omega/(2\pi\epsilon h_0 p_0) $ & $1$ to $10^3$ & -- \\
    \hline\hline
    \end{tabular}
    \caption{The dimensional and dimensionless parameters of the model for a two-dimensional channel with a compliant wall. The typical fluid is taken to be water, while the typical elastic solid is taken to be polydimethylsiloxane (PDMS), for which $\rho_s\simeq\rho_f$, $\nu_s \simeq 0.5$, and $E$ can be varied. The stiffness constant is estimated as $k = 0.272(1-\nu_s^2)h_0/E$ \cite{WC22}, while the deformation scale is calculated as $\mathcal{U} = \beta h_0$ (see Sec.~\ref{sec:reduced_model_channel}).}
    \label{table:param_2D}
\end{table}

Now, if the oscillatory pressure gradient along the channel were separable and time-harmonic, as $-\partial p/\partial z = \Real[G
\re^{\ri \omega t}]$, then, the post-transient oscillatory flow solution, $v_z(y,t) = \Real[f(y)\re^{\ri \omega t}]$, to Eq.~\eqref{eq:z_lubrication_2D} is easily found in complex form (leaving the `$\Real[\,\cdot\,]$' understood):
\begin{equation}
    v_z(y,t) = 
    \frac{h_0^2}{\mu_f}\frac{1}{\ri{\alpha}^2}\left[1-\frac{\cos\left(\ri^{3/2} (1-2y/h){\alpha}\mathfrak{h}/2\right)}{\cos\left(\ri^{3/2} {\alpha}\mathfrak{h}/2 \right)}\right] \underbrace{G \re^{\ri \omega t }}_{\equiv -{\partial p}/{\partial z}}.
    \label{eq:vz_osc_channel}
\end{equation}
See also \cite[p.~89]{LL87f} and the discussion in \cite{M08,TKR20} for several other forms of the solution. 
To introduce $\alpha$ into the solution~\eqref{eq:vz_osc_channel}, we let $h = h_0 \mathfrak{h}$.
From Eq.~\eqref{eq:vz_osc_channel}, the volumetric flow rate is found to be
\begin{equation}
    q := \int_0^w\!\!\int_{0}^{h=h_0\mathfrak{h}} v_z \,\mathrm{d}y\mathrm{d}x = 
    \frac{w h_0^3\mathfrak{h}}{\mu_f} \frac{1}{\ri \alpha^2}\left[1 - \frac{1}{\ri^{3/2}{\alpha}\mathfrak{h}/2}\tan\left(\ri^{3/2}{\alpha}\mathfrak{h}/2\right)\right]\left(-\frac{\partial p}{\partial z}\right),
    \label{eq:flow_rate_chan}
\end{equation}

For oscillatory flow in a rigid channel, the height is uniform ($\mathfrak{h}=1$), the pressure gradient is constant ($G=G_0$), and Eq.~\eqref{eq:flow_rate_chan} can be directly integrated as an ordinary differential equation to find the relationship between the amplitude of the flow rate's oscillations and the applied pressure gradient's constant amplitude. For oscillatory flow in a non-uniform (or deformable) channel, however, the dimensionless channel height $\mathfrak{h}$ and pressure gradient $G$ are not constant, so further closures are needed, which we now discuss.

\subsection{Model for the elastic deformation of the channel wall}
\label{sec:deformation_model_channel}
First, we must specify how the 
height of the channel varies. We are interested in the case of \emph{flow-induced} deformation. Therefore, the channel height varies with the applied load from the hydrodynamic pressure in the channel. This type of channel height variation can generally be expressed as (see, \textit{e.g.}, \cite{WC21,WC22,C21}):
\begin{equation}
     h(p) = h_0 + kp = h_0 \underbrace{(1 + kp/h_0)}_{\mathfrak{h}(p)},
     \label{eq:h_z_chan}
\end{equation}
where $k$ is an effective ``stiffness'' constant that can be related to the elastic properties of the compliant wall, as well as the geometry \cite{C21}. In the context of blood oxygenation in the lungs, Fung \cite[Sec.~6.8]{F97} uses Eq.~\eqref{eq:h_z_chan} to model the elasticity of the pulmonary alveolar sheet. The deformation--pressure relation implied by Eq.~\eqref{eq:h_z_chan}, namely $u_y := h-h_0 = kp$, can also be obtained from the reduced deformation model of a 3D microchannel proposed in \cite{WC22}:
\begin{equation}
    \underbrace{\rho_s b\frac{\partial^2 u_y}{\partial t^2}}_{\text{inertia},~\mathcal{O}(St_s)}+\underbrace{\frac{u_y}{k}}_{\text{stiffness},~\mathcal{O}(1)} - \underbrace{\chi_t \frac{\partial^2 u_y}{\partial z^2}}_{\text{tension},~\mathcal{O}(\theta_t)} + \underbrace{\chi_b \frac{\partial^4 u_y}{\partial z^4}}_{\text{bending},~\mathcal{O}(\theta_b)} = \underbrace{p}_{\text{load},~\mathcal{O}(1)},
    \label{eq:uy_2D}
\end{equation}
where $b$ is an (effective) thickness of the fluid--solid interface, which is the same as the thickness of the wall in our model, $k$ is the stiffness of the wall, $\chi_t$ is the tension per unit length, and $\chi_b$ is the plate-like bending rigidity. Equation~\eqref{eq:uy_2D} is also commonly used in models of high-speed flow over compliant coatings (the so-called ``Kramer's surface'') \cite{RGHM88}. 

Next, we denote the characteristic scale of $u_y$ as $\mathcal{U}$ and show it can be determined by balancing the surface stiffness term (the second term on the left) with the flow pressure. Let us also introduce $St_s = {\rho_s b \mathcal{U}}\omega^2/(4\pi^2 p_0 )$ as the solid's Strouhal number, which represents the (squared) ratio of a characteristic solid deformation time scale ($\sim \sqrt{\rho_s b \mathcal{U}/p_0}$) to the characteristic fluid flow time scale chosen earlier ($2\pi/\omega$) \cite{IWC20}. In the present analysis (as in \cite{WC22} but unlike \cite{IWC20}), we assume that the wall deformation develops faster than the flow so that $St_s\ll1$. Then, the solid inertia is a weak effect, and we neglect it at the leading order. We can also write $\chi_t = \Bar{E}b \varepsilon_z$ and $\chi_b = \Bar{E} b^3/12$, where $\Bar{E}=E/(1-\nu_s^2)$, with $E$ and $\nu_s$ being the Young's modulus and the Poisson's ratio of the solid wall, respectively, and $\varepsilon_z$ is the longitudinal strain resulting from either weak pretension or the bulging of the wall. As an example for the latter, for a von K\'arm\'an beam, $\varepsilon_z\sim (\mathcal{U}/\ell)^2$ is given in \cite{IWC20}. Assuming that $\mathcal{U}\ll\ell$ and also the compliant top wall is made slender with $b\ll\ell$, it is not difficult to show that $\theta_t = \chi_t\mathcal{U}/(p_0\ell^2) \sim (b/\ell)(\mathcal{U}/\ell)^3 \ll 1 $ and $\theta_b = \chi_b\mathcal{U}/(p_0 \ell^4)\sim (b/\ell)^3(\mathcal{U}/\ell)\ll 1$, so that the bending and tension are negligible. Then, from Eq.~\eqref{eq:uy_2D}, we obtain $u_y=kp$ at leading order, so that $\mathcal{U} = \beta h_0$. Here, $\beta:=kp_0/h_0$ is the dimensionless FSI number, which gauges the strength of fluid--solid coupling.

Note, however, that though Eq.~\eqref{eq:h_z_chan} may, on the face of it, appear to be a Winkler-foundation-like model \cite{DMKBF18} for deformation, no such assumption needs to be made here (see \cite{WC22}), unlike earlier works \cite{SM04,YK05}. For a ``truly'' 2D elastic wall, an \emph{incompressible} Winkler-foundation-like model has certain limitations, as discussed in \cite{CV20}. Having previously derived an ``effective'' 2D elastic model from a 3D one obviates this issue.

\subsection{Reduced model: Governing equation for the pressure}
\label{sec:reduced_model_channel}

Following the standard procedure (see, \textit{e.g.}, \cite{panton,P80,S17_LH}), the conservation of mass equation~\eqref{eq:com_2D} can be integrated over $y\in[0,h]$, and using the kinematic condition $v_y(y=h,t)={\partial h}/{\partial t}$ yields the continuity equation
\begin{equation}\label{eq:Mass_Conservation}
     \frac{\partial q}{\partial z}+\frac{\partial A}{\partial t}=0,
\end{equation}
where $A=wh(p)$ is the channel's cross-sectional area. Recognizing that we have `complexified' $q$ and $p$ in Eq.~\eqref{eq:flow_rate_chan}, we interpret the real-valued area to be $A=wh(\Real[p])$, consistent with the fact that Eq.~\eqref{eq:h_z_chan} is a quasi-static linear algebraic equation. Then, using Eq.~\eqref{eq:h_z_chan}, we interpret Eq.~\eqref{eq:Mass_Conservation} as
\begin{equation}\label{eq:Mass_Conservation_Re}
     \frac{\partial \Real[q]}{\partial z}+\frac{\partial\{w(h_0+k\Real[p])\}}{\partial t} =  \frac{\partial \Real[q]}{\partial z} + wk\frac{\partial\Real[p]}{\partial t}=0.
\end{equation}
Clearly, we can just solve this PDE for the complexified $q$ and $p$, taking the real part afterward.
Thus, on substituting $q$ from Eq.~\eqref{eq:flow_rate_chan} with $h(\Real[p])=h_0\mathfrak{h}(\Real[p])$ in the complexified Eq.~\eqref{eq:Mass_Conservation_Re}, we obtain a complex-valued, nonlinear PDE for the pressure:
\begin{equation}
       \frac{h_0^3}{\mu_f }\frac{\partial}{\partial z}\left\{-\frac{\partial p}{\partial z}\mathfrak{h}(\Real[p]) \frac{1}{\ri \alpha^2} \left[1-\frac{1}{\ri^{3/2}{\alpha}\mathfrak{h}(\Real[p])/2}\tan\left(\ri^{3/2}{\alpha}\mathfrak{h}(\Real[p])/2\right)\right]\right\}
       + k\frac{\partial p}{\partial t} = 0.
       \label{eq:p_pde_channel_dim}
\end{equation}

Next, we introduce dimensionless variables (based on the scales from Table~\ref{table:scales_2D}),  denote them by capital letters, and eliminate $\mathfrak{h}(\Real[p])$ via Eq.~\eqref{eq:h_z_chan} from Eq.~\eqref{eq:p_pde_channel_dim}, to obtain:
\begin{equation}
    \frac{\partial}{\partial Z}\left\{
    -\frac{\partial P}{\partial Z}\left(1+\beta \Real[P]\right)\frac{1}{\ri {\alpha}^2}\left[1-\frac{1}{\ri^{3/2}{\alpha}\left(1+\beta \Real[P]\right)/2}\tan\left(\ri^{3/2}{\alpha}\left(1+\beta \Real[P]\right)/2\right)\right]\right\}
    + \beta St_f \frac{\partial P}{\partial T}=0,
    \label{eq:p_pde_channel}
\end{equation}
where $\beta := k p_0/h_0$ has been defined as the FSI (or, compliance) number, and $St_f := (\ell/\mathcal{V}_z)/(2\pi/\omega) = \ell \mu_f \omega/(2\pi\epsilon h_0 p_0)$ is an axial Strouhal number for the flow. Equation~\eqref{eq:p_pde_channel} is a complex-valued, nonlinear PDE for the pressure distribution $P(Z,T)$ accounting for the oscillatory flow in the 2D channel two-way coupled to the flow-induced deformation of the channel's top elastic wall.  

The boundary conditions for Eq.~\eqref{eq:p_pde_channel} corresponding to time-harmonic oscillatory flow (driven by a pressure difference along the channel) are%
\begin{subequations}\label{eq:p_bc}
\begin{align}
    P(Z=0,T) &= \re^{2\pi \ri T},\\
    P(Z=1,T) &= 0.
\end{align}\end{subequations}
Although the PDE~\eqref{eq:p_pde_channel} is based on the long-time, post-transient flow solution, it still requires an initial condition on $P$ to be marched forward in time. For simplicity, we may impose a zero initial pressure distribution:
\begin{equation}\label{eq:p_ic}
    P(Z,T=0) = 0,
\end{equation}
and solve the PDE~\eqref{eq:p_pde_channel}  for ``sufficiently large'' $T$ so that this initial condition is ``forgotten'' and a post-transient state is achieved. Finally, we will take the real part of the computed numerical solution of Eq.~\eqref{eq:p_pde_channel} for plotting and analysis.

Note that the velocity profile used to obtain the flow rate and the corresponding PDE \eqref{eq:p_pde_channel} assumes that the pressure gradient is separable as $-{\partial p}/{\partial z}=G(z)\re^{\ri \omega t}$,  however, the same separation of variable ansatz cannot be used to solve the governing PDE \eqref{eq:p_pde_channel}, owing to its nonlinear nature from two-way coupling of the FSI. This apparent contradiction is resolved by understanding that we have essentially assumed a  velocity profile to ``close'' the cross-sectionally-averaged model. Borrowing the terminology from boundary-layer flows, this closure is generally referred to as the von K\'{a}rm\'{a}n--Pohlhausen approximation in a (weakly-)inertial flow in a channel with a deformable wall \cite{SWJ09,BBUGS18,IWC20,WC21,WC22}, wherein a steady parabolic profile is used to close the cross-sectionally-averaged momentum equation. (A similar closure problem arises in depth-averaged models of weakly-inertial thin film flows \cite[Ch.~6]{KRSV12}.) Here, we use the unidirectional oscillatory flow solution toward the same goal. 

Unfortunately, as is the case with the von K\'{a}rm\'{a}n--Pohlhausen approximation, our approximation can only be justified \textit{a posteriori}. In particular, in Sec.~\ref{sec:Results_Tube}, by comparing it to 3D direct numerical simulations. The only other alternative is to consider a weakly deformable conduit and expand the governing equations in $\beta\ll1$ \cite{CLMT05,CTGMHR06,AC20}, which is often referred to as the ``domain perturbation'' approach. However, as \citet{VD87} argues, the latter approach is expected to have a more limited range of accuracy. Furthermore, even though the resulting coupled system of the unsteady mass, momentum, and elasticity equations is linear and can be solved analytically at each order of $\beta$ using, \textit{e.g.}, Green's function methods \cite{CLMT05,CTGMHR06}, the analytical expressions are unwieldy and not of practical use. Meanwhile, our approximation of using the separable oscillatory flow profile to eliminate the momentum equation yields a closed-form reduced-order, two-way coupled FSI model~\eqref{eq:p_pde_channel} that does not assume $\beta\ll1$ \textit{a priori}.

\subsection{Weakly deformable channel}
\label{sec:weak_FSI_channel}

Following \cite{CLMT05,CTGMHR06,AC20}, we can seek a perturbation solution to Eq.~\eqref{eq:p_pde_channel} for weak FSI (\textit{i.e.}, $\beta\ll1$). To this end, let
\begin{equation}
    P(Z,T) = P_0(Z,T) + \beta P_1(Z,T) + \cdots.
    \label{eq:p_expansion_channel}
\end{equation}
Judiciously expanding the nonlinear term within the $Z$ derivative and substituting the expansion from Eq.~\eqref{eq:p_expansion_channel} into Eq.~\eqref{eq:p_pde_channel}, we obtain:
\begin{equation}
    \frac{\partial}{\partial Z}\left\{-\frac{\partial }{\partial Z}(P_0+\beta P_1)\Big(\mathfrak{f}_0({\alpha})+\beta \Real[P_0+\beta P_1] \mathfrak{f}_1({\alpha})\Big)\right\} + \beta St_f \frac{\partial (P_0+\beta P_1)}{\partial T} = 0,
    \label{eq:p_pde_expanded_channel}
\end{equation}
where, for convenience, we have defined
\begin{subequations}\begin{align}
    \mathfrak{f}_0({\alpha}) &:= \frac{1}{\ri \alpha^2} \left[1 - \frac{1}{\ri^{3/2}{\alpha}/2} \tan\big(\ri^{3/2}{\alpha}/2\big)\right],\\
    \mathfrak{f}_1({\alpha}) &:=  -\frac{1}{\ri \alpha^2} 
 \tan^2\big(\ri^{3/2}{\alpha}/2\big). 
\end{align}\end{subequations}
In passing, we note that $1/\mathfrak{f}_0(\alpha)$ is the contribution of the hydraulic resistance to the so-called hydraulic \emph{impedance} of a channel \cite{MF04,K10}.

Assuming that $\beta St_f = \mathcal{O}(\beta)$ asymptotically, collecting $\mathcal{O}(1)$ terms in Eq.~\eqref{eq:p_pde_expanded_channel} yields
\begin{equation}
    \frac{\partial^2 P_0}{\partial Z^2}=0
    \label{eq:p0_pde_channel}
\end{equation}
subject to 
\begin{subequations}\label{eq:p0_bcs_channel}
\begin{align}
    P_0(Z=0,T) &= \re^{2\pi \ri T}, \label{eq:p0_z0_bc_channel}\\
    P_0(Z=1,T) &= 0.
\end{align}\end{subequations}
The solution to the leading-order problem, \textit{i.e.},  Eqs.~\eqref{eq:p0_pde_channel} and \eqref{eq:p0_bcs_channel}, is thus simply:
\begin{equation}
    P_0(Z,T) = (1-Z) \re^{2\pi \ri T}.
    \label{eq:P0_chan}
\end{equation}
Note that, from Eqs.~\eqref{eq:p_pde_expanded_channel} and \eqref{eq:P0_chan}, we can easily deduce the known relation, $Q_0 = \Real[\mathfrak{f}_0(\alpha) \re^{2\pi \ri T}]$, between the rigid-channel flow rate $Q_0$ and the pressure drop (of unit value in our nondimensionalization scheme).

Next, collecting $\mathcal{O}(\beta)$ terms in Eq.~\eqref{eq:p_pde_expanded_channel} yields
\begin{equation}
    \mathfrak{f}_0({\alpha})\frac{\partial^2 P_1}{\partial Z^2}
    = -\frac{\partial}{\partial Z}\left[\mathfrak{f}_1({\alpha})\Real[P_0]\frac{\partial P_0}{\partial Z}\right] +  St_f \frac{\partial P_0}{\partial T}
    \label{eq:p1_pde_channel}
\end{equation}
subject to
\begin{subequations}\label{eq:p1_bcs_channel}
\begin{align}
    P_1(Z=0,T)&=0,\\
    P_1(Z=1,T)&=0.
\end{align}\end{subequations}
The solution for the first-order correction is found, from Eqs.~\eqref{eq:p1_pde_channel} and \eqref{eq:p1_bcs_channel},  to be:
\begin{equation}
    P_1(Z,T) = \frac{1}{6} Z(1-Z)\left[3\frac{\mathfrak{f}_1({\alpha})}{\mathfrak{f}_0({\alpha})} \Real[ \re^{2\pi \ri T}] \re^{2\pi \ri T} + (Z - 2)\frac{St_f}{\mathfrak{f}_0({\alpha})} 2\pi \ri \re^{2\pi \ri T} \right].
    \label{eq:P1_chan}
\end{equation}

Now, we define the cycle average as
\begin{equation}
    \langle \cdot \rangle(Z) := \int_{T}^{T+1} (\cdot)(Z,T') \, \rd T',
    \label{eq:cycle_avg}
\end{equation}
recalling that the dimensionless period is unity in our chosen dimensionless variables.
The perturbative, real-valued pressure distribution is then found from Eqs.~\eqref{eq:P0_chan} and \eqref{eq:P1_chan} as $\Real[P_0(Z,T)] + \beta \Real[P_1(Z,T)]$. 
Using the cycle averaging defined in Eq.~\eqref{eq:cycle_avg}, we find that the real-valued cycle-averaged pressure is
\begin{equation}\label{eq:P_avg_channel}
    \begin{aligned}
    \langle P \rangle(Z) = \Real[\underbrace{\langle P_0 \rangle}_{=0} + \beta\langle P_1 \rangle] + \mathcal{O}(\beta^2)
    &= \frac{\beta}{4} Z(1-Z) \Real\left[\frac{\mathfrak{f}_1({\alpha})}{\mathfrak{f}_0({\alpha})}\right] + \mathcal{O}(\beta^2)\\
    &= \frac{\beta}{4} Z(1-Z) \left(3 - \frac{43}{2100}{\alpha}^4\right) + \mathcal{O}(\beta{\alpha}^8,\beta^2),
    \end{aligned}
\end{equation}
where we have given the ${\alpha}\ll1$ expansion for completeness. The nested $\Real[\Real[\,\cdot\,]]$ expression arising in calculating the result in Eq.~\eqref{eq:P_avg_channel} leads to a product of the form $\Real[\,\cdot\,]\Real[\,\cdot\,]$, which can be evaluated using the identity $\Real[\mathfrak{w}_1]\Real[\mathfrak{w}_2]=\tfrac{1}{2}\{\Real[\mathfrak{w}_1\mathfrak{w}_2]+\Real[\mathfrak{w}_1\mathfrak{w}_2^*]\}$, for any $\mathfrak{w}_1,\mathfrak{w}_2\in\mathbb{C}$  \cite[p.~188]{BAH87}. For plotting, $\Real[\mathfrak{f}_1({\alpha})/\mathfrak{f}_0({\alpha})]$ in Eq.~\eqref{eq:P_avg_channel} is evaluated numerically.

Evidently, Eq.~\eqref{eq:P_avg_channel} implies the existence of a steaming pressure gradient ${\partial\langle P \rangle(Z)}/{\partial Z}$, 
which engenders a nonzero mean flow rate $\langle{Q}\rangle$. From the dimensionless flow rate corresponding to  Eq.~\eqref{eq:flow_rate_chan} a lengthy, but straightforward, calculation shows that
\begin{equation}
    \langle Q\rangle 
    = \frac{\beta}{4} \Real\left[\mathfrak{f}_1 (\alpha) \right] + \mathcal{O}(\beta^2)\\
    = \frac{\beta}{16}\left(1 - \frac{17}{720} \alpha^4 \right) + \mathcal{O}(\beta\alpha^6,\beta^2),
    \label{eq:Qavg_chan}
\end{equation}
which is independent of $Z$.
Further, $\langle Q\rangle/\beta$ is a decreasing function for small $\alpha$ and decays to zero as $\alpha\to\infty$. Interestingly, $\langle Q\rangle/\beta$ becomes negative between $\alpha \approx 4.4429$ and $\alpha \approx 8.8858$.

\subsection{Numerical results and discussion}\label{sec:results_chan}

The proposed reduced-order model, namely the PDE~\eqref{eq:p_pde_channel}, is a non-degenerate nonlinear diffusion equation. However, we need to solve this PDE numerically because, as discussed above, the pressure cannot be assumed to be time-harmonic (as it would be in the classical Womersley-style one-way coupled FSI analysis), and two-way coupling of the flow and deformation leads to a nonlinear PDE. There are many numerical methods suitable for solving such a PDE with ease \cite{SB90}. For convenience, we simply use the built-in \texttt{pdepe} of \textsc{Matlab} 2020b (Mathworks, Inc.) to solve Eq.~\eqref{eq:p_pde_channel} subject to Eqs.~\eqref{eq:p_bc} and \eqref{eq:p_ic}. \texttt{Pdepe} uses an auto-generated finite-element spatial discretization of the nonlinear parabolic (or elliptic) PDE provided \cite{SB90} and the method of lines for time integration, which is accomplished by \textsc{Matlab}'s adaptive, variable-order multistep stiff solver \texttt{ode15s} \citep{SR97}. The relative tolerance of the solver is set to $10^{-15}$, while the absolute tolerance is set to $10^{-8}$. A total of $1000$ spatial grid points were used, having verified grid convergence. Although Eq.~\eqref{eq:p_pde_channel} is a \emph{complex-valued} PDE, it depends only upon the real variables $Z$ and $T$, hence it can be solved using \texttt{pdepe} just like a real-valued PDE.

\begin{figure}
    \centering
    \begin{subfigure}[b]{0.49\textwidth}
         \centering
         \includegraphics[width=\textwidth]{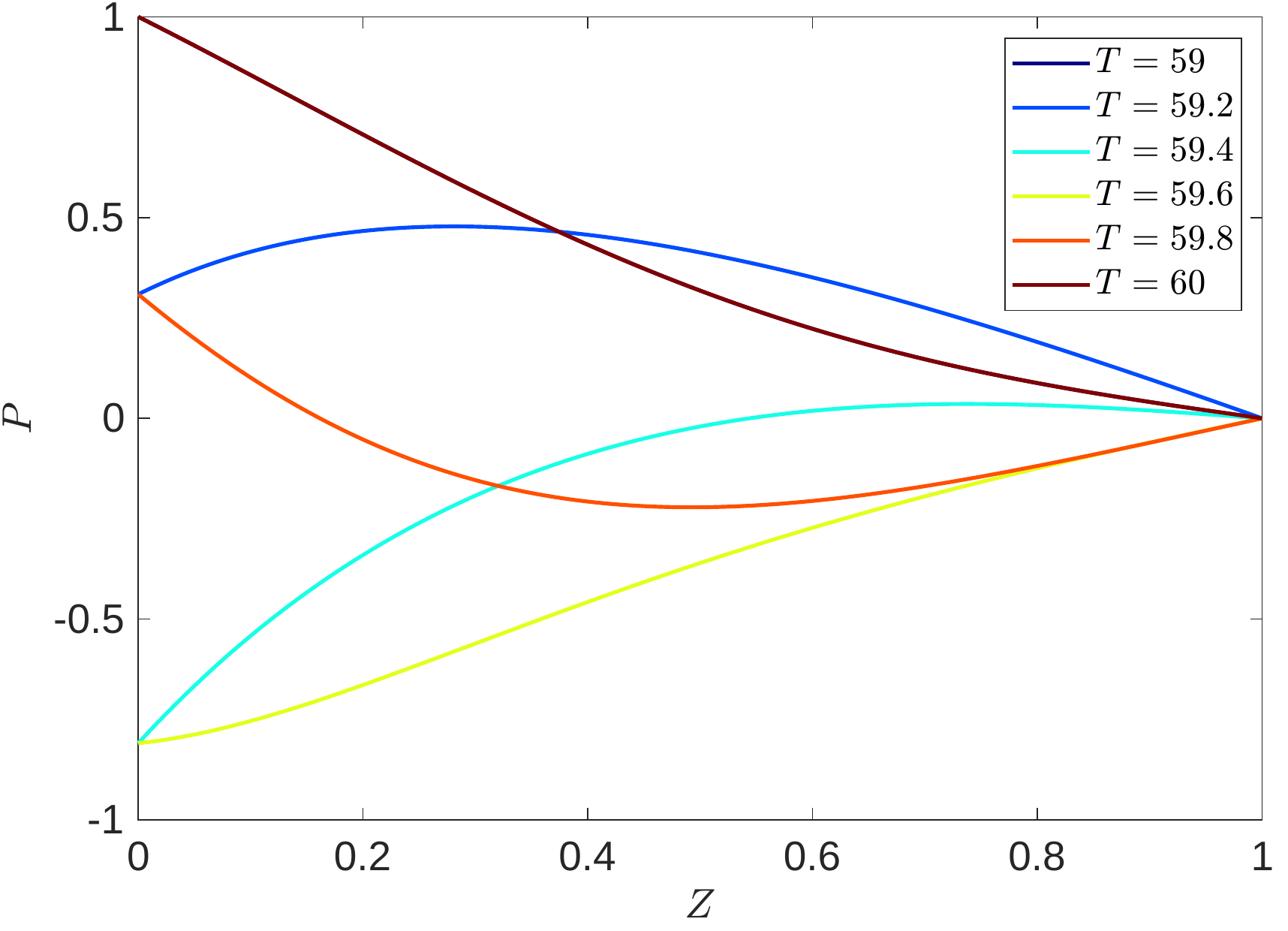}
         \caption{}
         \label{fig:pressure_channel}
     \end{subfigure}
     \label{pdepe_channel}
     \hfill
     \begin{subfigure}[b]{0.49\textwidth}
         \centering
         \includegraphics[width=\textwidth]{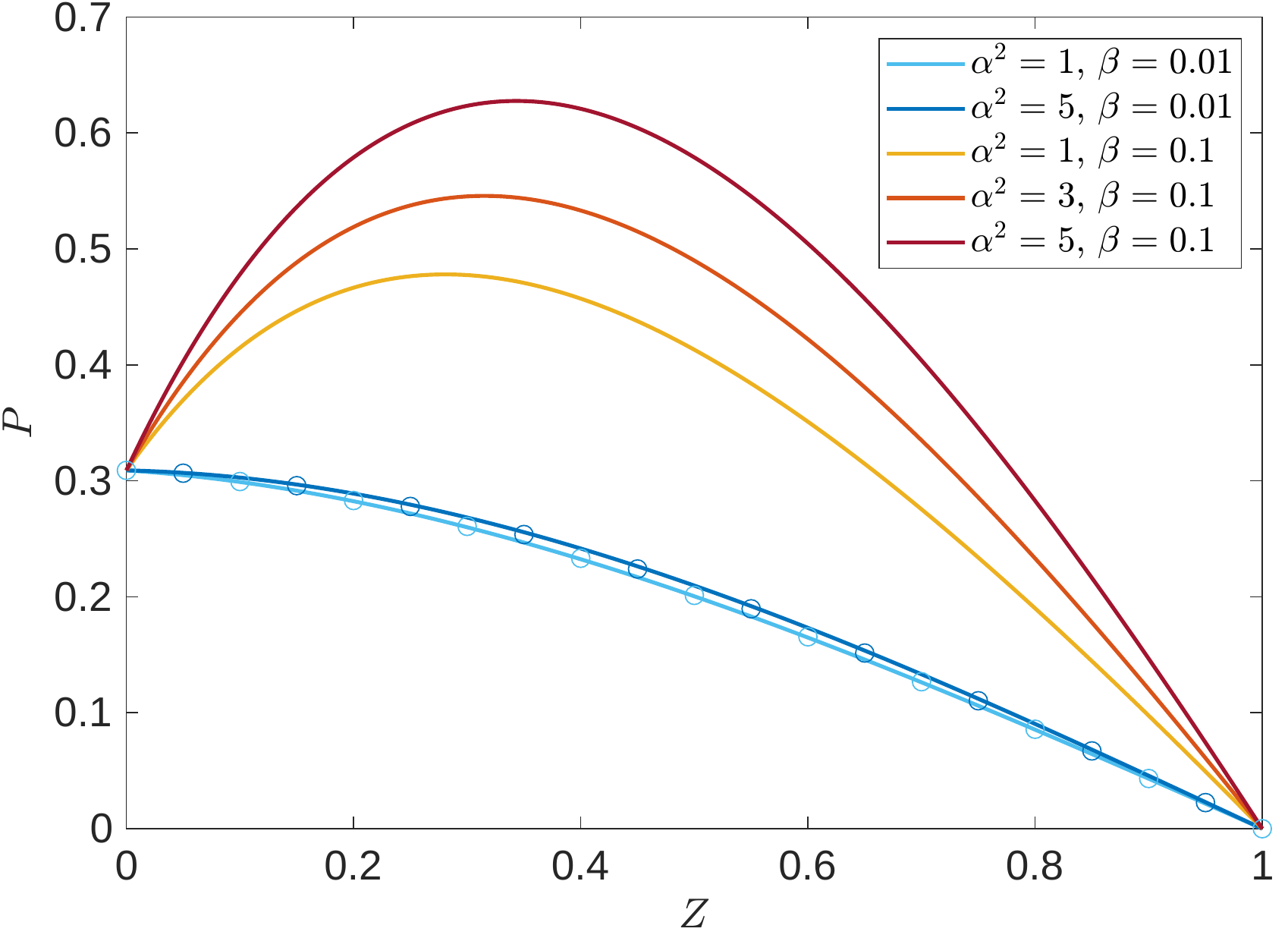}
         \caption{}
         \label{fig:distribution_p_channel}
     \end{subfigure}
    \label{fig:P_channel}
     \begin{subfigure}[b]{0.49\textwidth}
         \centering
         \includegraphics[width=\textwidth]{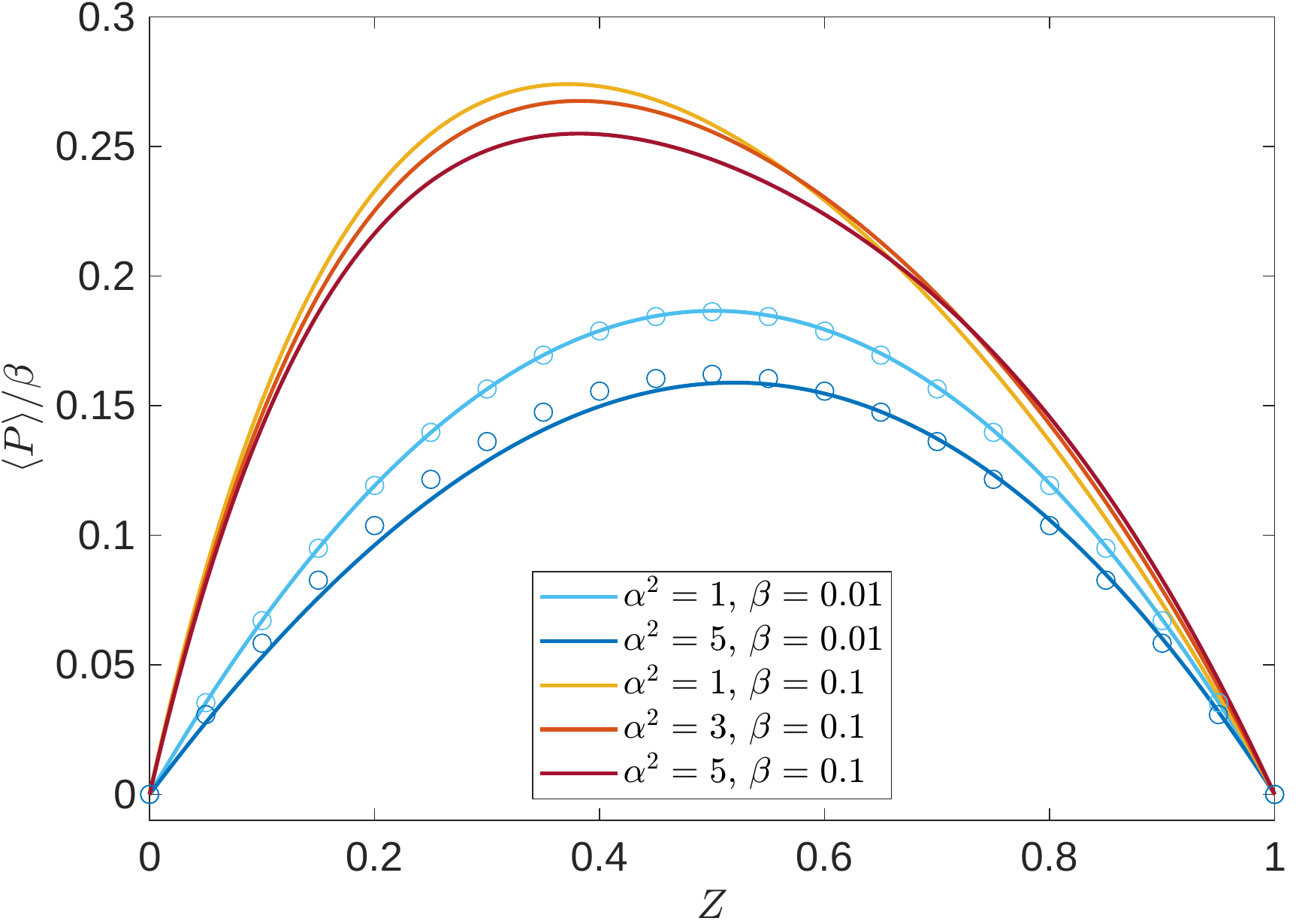}
         \caption{}
         \label{fig:streaming_p_channel}
     \end{subfigure}
    \label{fig:P_Dist}
    \caption{(a) The dimensionless pressure distribution at different times, computed by numerically solving Eq.~\eqref{eq:p_pde_channel} subject to Eqs.~\eqref{eq:p_ic} and \eqref{eq:p_bc} for ${\alpha}^2=1$, $\beta=0.1$, and $St_f=1$. (b) The dimensionless pressure distribution from (a) at $T=59.2$ for different values of the Womersley number $\alpha$ and the compliance number $\beta$. Symbols ($\circ$) represent the analytical perturbation solution $\Real[P_0+\beta P_1]$ found from Eqs.~\eqref{eq:P0_chan} and \eqref{eq:P1_chan}. (c) The corresponding ``universal'' normalized streaming pressure $\langle P \rangle/\beta$ profiles. Symbols ($\circ$) represent the analytical perturbation solution found from Eq.~\eqref{eq:P_avg_channel}.}
\end{figure}

An example numerical solution is shown in Fig.~\ref{fig:pressure_channel} over one cycle of the forcing, after sufficient time has elapsed for the solution to reach a time-periodic state (observe that the $P$ curves at the beginning and end of the cycle shown overlap). By ``sufficient time," we mean that the maximum pressure difference between two consecutive cycles is less than a prescribed tolerance, namely a $T$ such that $\max_Z|P(Z,T)-P(Z,T+1)|<10^{-8}$. 

Having established a solution procedure for the governing PDE, in Fig.~\ref{fig:distribution_p_channel}, we next highlight how the pressure distribution obtained from solving Eq.~\eqref{eq:p_pde_channel} numerically varies with the Womersley number, and how it compares to the perturbation solution (for the cases of $\alpha^2=1$ and $\alpha^2=5$ and  $\beta=10^{-2}$). In Fig.~\ref{fig:streaming_p_channel}, we show the normalized cycled-averaged pressure $\langle P \rangle /\beta$, for the same parameters. The cycle averaged pressure for the numerical solution is calculated using the \texttt{trapz} function (trapezoidal rule for integration) in \textsc{Matlab}, with a dimensionless time-step of $\Delta T = 0.1$, while the cycle averaged pressure from perturbation solution is given by Eq.~\eqref{eq:P_avg_channel}. We verified that $\Delta T = 0.1$ is sufficient for the time-averaging (decreasing this $\Delta T$ does not change the averaged result). In  Fig.~\ref{fig:distribution_p_channel}, the perturbation solutions and the numerical solutions are in good agreement with each other (maximum difference of $<1\%$), thus demonstrating the utility of the analytical results from Sec.~\ref{sec:weak_FSI_channel} in the weak FSI regime ($\beta\ll1$).

Observe that $\langle P \rangle/\beta$, as calculated in Eq.~\eqref{eq:P_avg_channel} using the perturbation solution, has a ``universal'' shape with respect to $Z$, in the sense it only depends solely upon $\alpha$ and no further details of the FSI. Figure~\ref{fig:streaming_p_channel} shows that the perturbative result agrees very well with the numerical solution (maximum difference of $<5\%$) for different Womersley numbers, demonstrating that FSI leads to a nonzero cycle-averaged pressure distribution, despite the inlet forcing having a zero mean. For larger values of the compliance number $\beta$, the cycle-averaged pressure from the simulations is not only ``stronger'' (larger values) but also becomes asymmetric, with its maximum shifting towards the inlet.

Riley \cite{R01} defines \emph{steady streaming} to refer to precisely the latter phenomenon, namely when the ``time-average of a fluctuating flow often results in a nonzero mean.'' More generally, viscous streaming refers to the induction of a steady mean flow from time-harmonic oscillations (of the boundaries, inlet conditions, or another mechanism driving the flow). One of the most well-known examples of streaming arises due to small-amplitude, high-frequency oscillations of a body in a viscous (or inviscid) fluid \cite{Riley98}, as famously featured in Van Dyke's \textit{An Album of Fluid Motion} \cite[p.~23]{VD_Album}. Although classically streaming is induced by the motion of \emph{rigid} objects (or boundaries) in a flow, \emph{soft} streaming has become of interest recently in the context of both external \cite{BPG22} and internal \cite{AC20} flows. 
In the context of the present problem of flow in a slender conduit, another classical example of streaming in a viscous flow in a channel is the so-called mechanism of \emph{peristaltic pumping}, which \citet{JS71} define as ``fluid transport that occurs when a progressive wave of area contraction or expansion propagates along the length of a distensible tube containing a liquid.'' Traditionally, however, the viscous flow in peristalsis is driven by moving a wavy wall, but \citet{FY68} have speculated that peristalsis may be related to the spontaneous oscillations of blood vessels (``vasomotion''), perhaps somewhat akin to the present context in which the conduit walls are not externally actuated.

Earlier work by Hall \cite{H74} showed that weak inertia (at the leading order in a suitable Reynolds number) generates a streaming flow when an oscillatory pressure difference is maintained between the ends of a tube of axially varying radius. This phenomenon was successfully analyzed by perturbation expansions for both small and large Womersley numbers \cite{H74}. Recently, however, it was further demonstrated that viscous streaming can arise even a vanishing Reynolds number if the tube radius' axial variations are due to two-way coupled FSI with the flow \cite{AC20}. In a sense, FSI self-generates peristaltic pumping without the need for external intervention (such as moving the wall). However, this zero-Reynolds-number mechanism was only analyzed in \cite{AC20} at low Womersley numbers and for compressible flow. On the other hand, as our results in this section demonstrate, the model proposed in this work is able to capture viscous streaming induced by FSI at an \emph{arbitrary} Womersley number in an incompressible flow.


\section{Oscillatory flow in an axisymmetric compliant tube}
\label{sec:tube}

\subsection{Governing equations: scaling and lubrication approximation}
\label{sec:lubrication_tube}

\begin{figure}
    \centering
    \includegraphics[width=0.6\textwidth]{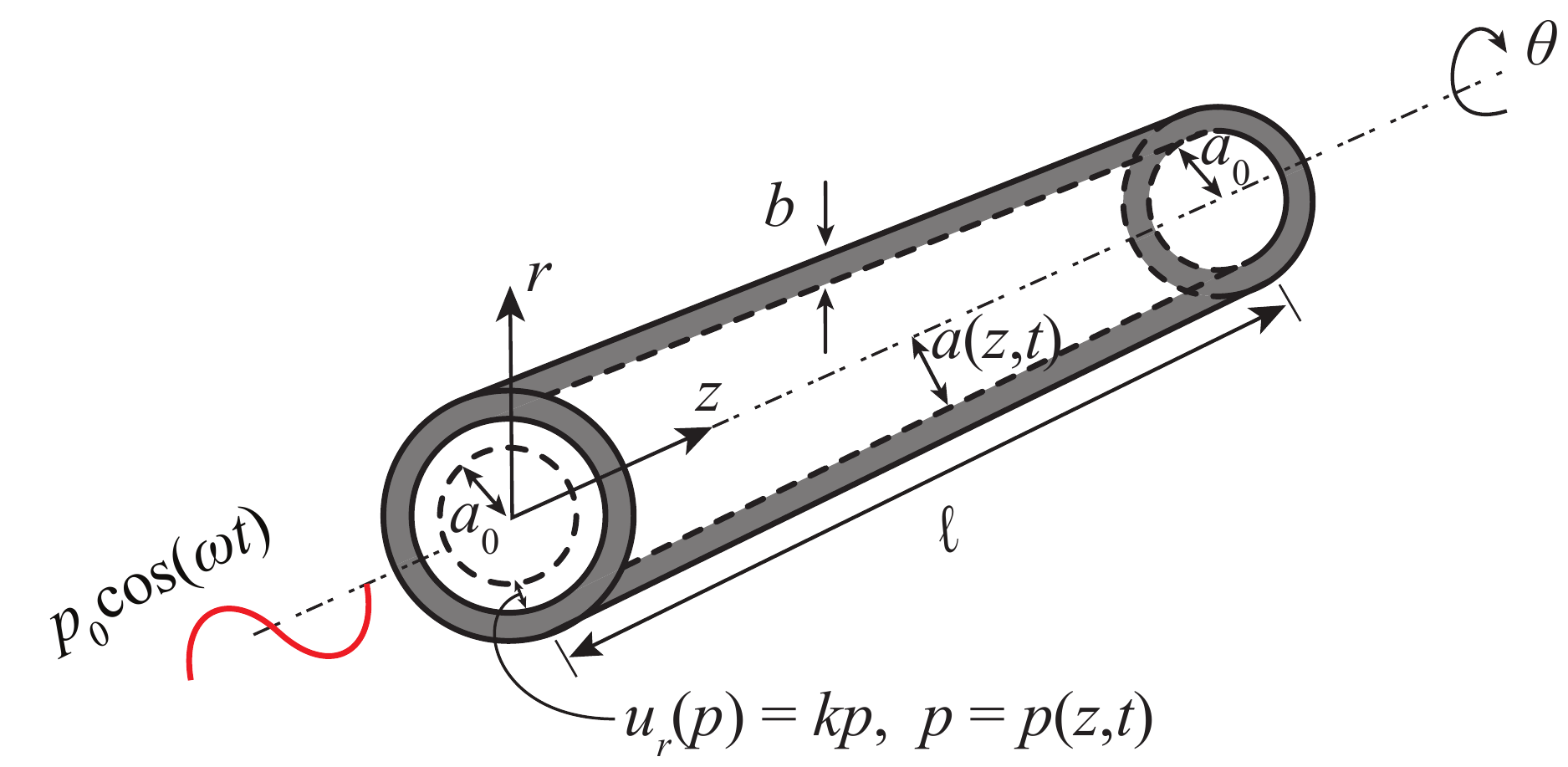}
    \caption{Schematic of an axisymmetric compliant microtube, indicating key quantities and notation for the geometry.}
    \label{fig:schematic_tube}
\end{figure}

Consider a pressure-driven axisymmetric flow without swirl, such that $v_\theta=0$ and ${\partial (\cdot)}/{\partial \theta}=0$, of a Newtonian fluid in a cylindrical tube with $z$ being the axial direction, as shown in Fig.~\ref{fig:schematic_tube}. Then, neglecting body forces, the mass and momentum conservation equations for this flow \cite{panton} are
\begin{subequations}
\begin{align}
\label{eq:com_AS}
   \underbrace{\frac{1}{r}\frac{\partial}{\partial r}(rv_r)}_{\mathcal{O}(1)}+\underbrace{\frac{\partial v_z}{\partial z}}_{\mathcal{O}(1)} &= 0,\\
\label{eq:r_momentum_AS}
    \underbrace{\rho_f\frac{\partial v_r}{\partial t}}_{\mathcal{O}(\epsilon^2{\alpha}^2)}
    +\underbrace{\rho_f v_r\frac{\partial v_r}{\partial r}}_{\mathcal{O}(\epsilon^3Re)}+\underbrace{\rho_f v_z\frac{\partial v_r}{\partial z}}_{\mathcal{O}(\epsilon^3 Re)} &= \underbrace{\mu_f\frac{\partial}{\partial r}\left[\frac{1}{r}\frac{\partial}{\partial r}(r v_r)\right]}_{\mathcal{O}(\epsilon^2)}+\underbrace{\mu_f\frac{\partial^2 v_r}{\partial z^2}}_{\mathcal{O}(\epsilon^4)}-\underbrace{\frac{\partial p}{\partial r}}_{\mathcal{O}(1)},\\
\label{eq:z_momentum_AS}
    \underbrace{\rho_{f}\frac{\partial v_z}{\partial t}}_{\mathcal{O}({\alpha}^2)}   
   +\underbrace{\rho_f v_r\frac{\partial v_z}{\partial r}}_{\mathcal{O}(\epsilon Re)}+\underbrace{\rho_f v_z\frac{\partial v_z}{\partial z}}_{\mathcal{O}(\epsilon Re)} &= \underbrace{\mu_f\frac{1}{r}\frac{\partial}{\partial r}\left(r\frac{\partial v_z}{\partial r}\right)}_{\mathcal{O}(1)}+\underbrace{\mu_f\frac{\partial^2 v_z}{\partial z^2}}_{\mathcal{O}(\epsilon^2)}-\underbrace{\frac{\partial p}{\partial z}}_{\mathcal{O}(1)}.
\end{align}\label{eq:iNS_AS}\end{subequations}
The scales used to determine the orders of magnitude of terms in Eqs.~\eqref{eq:iNS_AS} are given in Table~\ref{table:scales_AS}. Three dimensionless numbers ($\epsilon$, $Re$ and $\alpha$) arise. They are defined in Table~\ref{table:param_AS}, where their typical values are also given.

\begin{table}[hb]
\begin{tabular}{l@{\qquad} l@{\qquad} l@{\qquad} l@{\qquad} l@{\qquad} l@{\qquad} l}
    \hline\hline
    Variable & $t$ & $r$ & $z$ & $v_r$ & $v_z$ & $p$\\
    \hline
    Scale & ${2 \pi}/{\omega}$ & $a_0$ & $\ell$ & $\epsilon \mathcal{V}_z$ & $\mathcal{V}_z = {\epsilon a_0 p_0}/{\mu_f}$ & $p_0$\\
    \hline\hline
    \end{tabular}
\caption{The scales for the variables in the axisymmetric incompressible Navier--Stokes equations~\eqref{eq:iNS_AS}. We have used the lubrication-theory scales for $v_r$, $v_z$ based on the pressure scale imposed by $p_0$.}
\label{table:scales_AS}
\end{table}

\subsection{Flow solution at arbitrary Womersley number}
\label{sec:flow_solution_tube}

Based on the typical values given in Table~\ref{table:param_AS}, we are led to consider the regime of $\epsilon\ll 1$, $\epsilon Re\ll 1$, $\epsilon^2{\alpha}^2\ll1$, which reduces Eq.~\eqref{eq:z_momentum_AS} to
\begin{equation}
    \rho_{f}\frac{\partial v_z}{\partial t}=\mu_f\frac{1}{r}\frac{\partial}{\partial r}\left(r\frac{\partial v_z}{\partial r}\right)-\frac{\partial p}{\partial z},
\label{eq:z_lubrication_AS}
\end{equation}
subject to no slip along the tube wall, $v_z(r=a,t)=0$. As before, Eq.~\eqref{eq:z_lubrication_AS} is valid for arbitrary ${\alpha}^2$ (as long as $\epsilon^2{\alpha}^2\ll1$).  As expected, Eq.~\eqref{eq:z_lubrication_AS} is the same in a rigid tube of constant radius $a_0$, as well as a deformable tube of variable radius $a$. Indeed, the tube radius $a$ does not have to be constant under the lubrication approximation, as long as it \emph{varies slowly} \cite{VD87}. (In this context, this result was also shown explicitly by  \citet{H74}.)

\begin{table}
    \centering
    \begin{tabular}{l@{\quad} l@{\quad} l@{\quad} l}
    \hline\hline
    Quantity & Notation & Typical value & Units \\
    \hline
     Tube's length & $\ell$ & $1$ to $100$ & \si{\milli\meter} \\
    Tube's undeformed radius & $a_{0}$ & $0.08$ to $0.5$ & \si{\milli\meter} \\
    Tube's thickness & $b$ & $0.008$ to $0.05$ & \si{\milli\meter} \\
    Solid's Young's modulus & $E$ &$0.5$ & \si{\mega\pascal}  \\
    Solid's Poisson's ratio & $\nu_s$ & $0.49$ to $0.5$ & -- \\
    Solid's density & $\rho_s$ & $1.0 \times 10^3$ & \si{\kilo\gram\per\meter\tothe{3}} \\
    Fluid's density & $\rho_f$ &$1.0\times10^{3}$ & \si{\kilo\gram\per\meter\tothe{3}} \\
    Fluid's dynamic viscosity & $\mu_f$ & $1.0\times10^{-3}$ & \si{\pascal\second} \\
    Pressure pulse amplitude & $p_0$ & $0.1$ to $2$ & \si{\kilo\pascal} \\
    Pressure pulse frequency & $\omega/ 2\pi$ & $1$ to $100$ & Hz \\
    \hline
    Tube's radius-to-length aspect ratio & $\epsilon={a_0}/{\ell}$ &$0.002$ to $0.06$ &  --\\
    Reynolds number & $Re={\rho_f \epsilon a_0^2 p_0}/{\mu_f^2}$  & $2$ to $400$ &  --\\
    Womersley number & ${\alpha}=a_0\sqrt{\rho_f \omega/\mu_f}$  & $0.3$  to $ 12$ &  --\\
    FSI (or, compliance) number & $\beta=k p_0/a_0 $ &  $0.005$ to $0.1$ & -- \\
    Solid's Strouhal number & $St_s = {\rho_s b \mathcal{U}\omega^2}/(4\pi^2 p_0)$ & 
    $\approx 10^{-10}$ to $10^{-8}$ & -- \\
    Fluid's Strouhal number & $St_f = \ell \mu_f \omega/(2\pi\epsilon a_0 p_0)$ & 1 to $10^3$ & -- \\
    \hline\hline
    \end{tabular}
    \caption{The dimensional and dimensionless parameters of the model for a compliant microtube. The typical fluid is taken to be water, while the typical elastic solid is taken to be PDMS, for which $\rho_s\simeq\rho_f$, $\nu_s \simeq 0.5$, and $E$ can be varied. The stiffness constant's order of magnitude is estimated as $k = (1-\nu_s^2){a_0^2}/(E b)$ \cite{AC18b}, while the deformation scale is calculated as $\mathcal{U} = \beta a_0$ (see Sec.~\ref{sec:reduced_model_tube}).}  
    \label{table:param_AS}
\end{table}

Now, if the oscillatory pressure gradient along the channel is separable and time-harmonic, as $-\partial p/\partial z = \Real[G
\re^{\ri \omega t}]$, then, the post-transient oscillatory flow solution, $v_z(r,t)=\Real[g(r)\re^{\ri\omega t}]$,  to Eq.~\eqref{eq:z_lubrication_AS} is known explicitly from, \textit{e.g.}, \citet{A16} (credited to Womersley, but also derived earlier by Sexl and contemporaneously by Uchida, see \cite[Sec.~4.6]{Z00}) in complex form (leaving the `$\Real[\,\cdot\,]$' understood):
\begin{equation}
    v_z(r,t) = \frac{a_0^2}{\mu_f} \frac{1}{\ri \alpha^2}\left[1-\frac{J_0\left(\ri ^{3/2}{\alpha} \mathfrak{a} r/a\right)}{J_0\left(\ri^{3/2}{\alpha}\mathfrak{a}\right)}\right]\underbrace{ G\re^{\ri\omega t}}_{\equiv -\partial p/\partial z},
    \label{eq:vz_osc_tube}
\end{equation}
where $J_n(\cdot)$ is the Bessel function of the first kind of order $n$. To introduce $\alpha$ into the solution~\eqref{eq:vz_osc_tube}, we let $a = a_0 \mathfrak{a}$.
As in the 2D case, the solution is given in complex-variable form for convenience, and we may take the real or imaginary part, depending on the boundary conditions. Then, from Eq.~\eqref{eq:vz_osc_tube} the flow rate--pressure gradient relation is found to be
\begin{equation}
    q(z,t) := 2\pi\int_0^{a=a_0\mathfrak{a}} v_z \,r \rd r = \frac{\pi a_0^4\mathfrak{a}^2}{\mu_f} \frac{1}{\ri \alpha^2} \left[1-\frac{2J_1\left(\ri^{3/2} {\alpha}\mathfrak{a}\right)}{\ri^{3/2} {\alpha}\mathfrak{a} J_0\left(\ri^{3/2} {\alpha}\mathfrak{a}\right)}\right]  \left(-\frac{\partial p}{\partial z}\right).
\label{eq:Worms_flow_rate}
\end{equation}

For oscillatory flow in a rigid tube, $\mathfrak{a}=1$, $G=G_0$, and Eq.~\eqref{eq:Worms_flow_rate} can be directly integrated as an ordinary differential equation to find the relationship between the amplitude of the flow rate's oscillations and the applied pressure gradient's constant amplitude. However, for oscillatory flow in a non-uniform (or deformable) tube, the dimensionless radius $\mathfrak{a}$ and the pressure gradient $G$ are not constant, so further closures are needed, which we now discuss.

\subsection{Model for the elastic deformation of the tube wall}
\label{sec:deformation_model_tube}
As in Sec.~\ref{sec:deformation_model_channel}, the variation of the radius of the tube, due to flow-induced deformation, can generally be expressed as (see, \textit{e.g.}, \cite{EG14,AC18b,C21}):
\begin{equation}\label{eq:a_z_tube}
    a(p) = a_0 + k p = a_0 \underbrace{(1 + k p/a_0)}_{\mathfrak{a}(p)},
\end{equation}
where $k$ is again an effective stiffness constant related to the elastic properties of the compliant wall, as well as its geometry. In the biofluid mechanics context, Eq.~\eqref{eq:a_z_tube} is often termed a ``tube law'' \cite{S77,A16}. The deformation--pressure relationship, $u_r := a - a_0 = kp$, implied by Eq.~\eqref{eq:a_z_tube} can be obtained from a suitable shell theory for thin cylindrical structures (see, \textit{e.g.}, \cite{LS88}). Assuming a long, slender axisymmetric tube, it has been argued in the biomechanics literature (see, \textit{e.g.}, \cite{CTGMHR06} and the numerous references therein and thereof) that linear shell theory generally yields an equation of motion of the form  
\begin{equation}\label{eq:Koiter Shell}
    \underbrace{\rho_s b\frac{\partial^2 u_r}{\partial t^2}}_{\text{inertia},~\mathcal{O}(St_s)} + \underbrace{\frac{u_r}{k}}_{\text{stiffness},~\mathcal{O}(1)} - \underbrace{\chi_t \frac{\partial^2 u_r}{\partial z^2}}_{\text{tension},~\mathcal{O}(\theta_t)} + \underbrace{\chi_b \frac{\partial ^4 u_r}{\partial z^4}}_{\text{bending},~\mathcal{O}(\theta_b)} = \underbrace{p}_{\text{load},~\mathcal{O}(1)}
\end{equation}
for the radial displacement $u_r$, having neglected axial displacements. Here, $St_s = {\rho_s b \mathcal{U}\omega^2}/(4\pi^2 p_0)$ is a solid's Strouhal number, and the scale for $u_r$ is again $\mathcal{U}$. 

Next, in the present analysis (as in \cite{EG14,EG16,BBG17} but unlike \cite{AC20}), we assume that the wall deformation develops faster than the flow so that $St_s\ll1$. Then,  the solid inertia is a weak effect, and we neglect it at the leading order. As an example, if Eq.~\eqref{eq:Koiter Shell} is derived from linear Koiter shell theory, we have $\chi_t = \Bar{E} b^3 \nu_s/(6a_0^2)$ and $\chi_b = \bar{E} b^3/12$ (see, \textit{e.g.}, \cite{CTGMHR06}).
The use of shell theory requires small strains ($\mathcal{U}\ll\ell$) and a thin ($b\ll a_0)$ and slender ($b\ll\ell$) tube. It follows that  $\theta_t = \chi_t\mathcal{U}/(p_0\ell^2) \sim (b/\ell)(b/a_0)^2(\mathcal{U}/\ell) \ll 1 $ and $\theta_b = \chi_b\mathcal{U}/(p_0\ell^4)\sim (b/\ell)^3(\mathcal{U}/\ell)\ll 1$, so that the bending and tension are negligible. Then, from Eq.~\eqref{eq:Koiter Shell} we obtain $u_r = kp$ at the leading order, so that $\mathcal{U}=\beta a_0$. (For alternative approaches, starting from the equations of linear elasticity and considering different geometric configurations, boundary conditions, and external loading, see \cite{EG14,EG16,WPC22}.)

\subsection{Reduced model: Governing equation for the pressure}
\label{sec:reduced_model_tube}

Following the standard procedure (see, \textit{e.g.}, \cite{panton,P80}), the conservation of mass equation~\eqref{eq:com_AS} for incompressible flow in an axisymmetric deforming tube can be shown to take the same integrated form as the continuity equation for a 2D channel, \textit{i.e.}, Eq.~\eqref{eq:Mass_Conservation} (see also \cite{GHJG96,PV09}). Then, following the same logic as in Sec.~\ref{sec:reduced_model_channel}, we obtain a complex-valued, nonlinear PDE for the pressure:
\begin{equation}\label{eq:pressure_dim}
    \frac{\pi a_0^4}{\mu_f} \frac{1}{\ri \alpha^2} \frac{\partial}{\partial z} \left\{-\frac{\partial p}{\partial z} \mathfrak{a}(\Real[p])^2 \left[1-\frac{2J_1\left(\ri^{3/2} {\alpha}\mathfrak{a}(\Real[p])\right)}{\ri^{3/2} {\alpha}\mathfrak{a}(\Real[p]) J_0\left(\ri^{3/2} {\alpha}\mathfrak{a}(\Real[p])\right)}\right]\right\} + 2\pi (a_0+k\Real[p])k\frac{\partial p}{\partial t} = 0.
\end{equation}

Next, we introduce dimensionless variables (based on the scales from Table~\ref{table:scales_AS}),  denote them by capital letters, and eliminate $\mathfrak{a}(\Real[p])$ via Eq.~\eqref{eq:a_z_tube} from Eq.~\eqref{eq:pressure_dim}, to obtain:
\begin{equation}
    \frac{\partial }{\partial Z} \left\{ - \frac{\partial P }{\partial Z}\left(1+ \beta \Real[P]\right)^2 \frac{1}{\ri \alpha^2} \left[1-\frac{2J_1\left(\ri^{3/2} {\alpha} \left(1+\beta \Real[P]\right)\right)}{\ri^{3/2} {\alpha} \left(1+\beta \Real[P]\right) J_0\left(\ri^{3/2} {\alpha} \left(1+\beta \Real[P]\right)\right)}\right]\right\} + {2 \beta St_f}(1+\beta\Real[P]) \frac{\partial P}{\partial T}= 0,
    \label{eq:dimless_P_pde}
\end{equation}
where $\beta := k p_0/a_0$ has been defined as the FSI (or, compliance) number, and $St_f := (\ell/\mathcal{V}_z)/(2\pi/\omega) = \ell \mu_f \omega/(2\pi\epsilon a_0 p_0)$ is an axial Strouhal number for the flow. As in Sec.~\ref{sec:reduced_model_channel}, for time-harmonic pressure-driven oscillatory flow, the dimensionless initial and boundary conditions for Eq.~\eqref{eq:dimless_P_pde} are once again given by Eqs.~\eqref{eq:p_ic} and \eqref{eq:p_bc}, respectively. As before, having left the `$\Real[\,\cdot\,]$' understood so far (except in the pressure--radius relation $\mathfrak{a}(p)$), below we will take the real part of the computed numerical solution of Eq.~\eqref{eq:dimless_P_pde} for plotting and analysis.

Again, we observe that although a separable form of the pressure gradient, in terms of a function of $z$ times a function of $t$, was used to obtain the flow profile~\eqref{eq:vz_osc_tube} from the reduced momentum equation~\eqref{eq:z_lubrication_AS} and close the relation~\eqref{eq:Worms_flow_rate} between flow rate and pressure gradient, the final PDE~\eqref{eq:dimless_P_pde} for $P(Z,T)$ is nonlinear and, strictly speaking, has no separable solutions. We reconcile this apparent contradiction, as done at the end of Sec.~\ref{sec:reduced_model_channel}, by noting that we have essentially used a von K\'{a}rm\'{a}n--Pohlhausen-type approximation to close the cross-sectionally-averaged model for oscillatory flow in a deformable conduit. This approximation's validity is checked \textit{a posteriori} in Sec.~\ref{sec:Results_Tube} by comparing Eq.~\eqref{eq:dimless_P_pde}'s predictions to 3D direct numerical simulations.

\subsection{Weakly deformable tube}
\label{sec:weak_FSI_tube}
As in Sec.~\ref{sec:weak_FSI_channel}, following \cite{CLMT05,CTGMHR06,AC20}, let us seek a perturbation solution for weak FSI (\textit{i.e.}, $\beta\ll1$). Judiciously expanding the nonlinear term within the $Z$ derivative and substituting the expansion from Eq.~\eqref{eq:p_expansion_channel} into Eq.~\eqref{eq:dimless_P_pde}, we obtain:
\begin{equation}
    \frac{\partial}{\partial Z}\left\{-\frac{\partial }{\partial Z}(P_0+\beta P_1)\Big(\mathfrak{g}_0({\alpha})+\beta\Real[P_0+\beta P_1]\mathfrak{g}_1({\alpha})\Big)\right\} + {2 \beta St_f} (1+\beta \Real[P_0+\beta P_1]) \frac{\partial }{\partial T} (P_0+\beta P_1) = 0,
    \label{eq:p_pde_expanded_tube}
\end{equation}
where, for convenience, we have defined
\begin{subequations}\begin{align}
    \mathfrak{g}_0({\alpha}) &:= \frac{1}{\ri \alpha^2} \left[ 1-\frac{2J_1\left(\ri^{3/2} {\alpha}\right)}{\ri^{3/2} {\alpha} J_0\left(\ri^{3/2} \alpha \right)} \right] = -\frac{J_2\left(\ri^{3/2} \alpha \right)}{\ri \alpha^2 J_0\left(\ri^{3/2} \alpha \right)},\\
    \mathfrak{g}_1({\alpha}) &:= -\frac{2 J_1\left(\ri^{3/2} \alpha \right)^2}{\ri \alpha^2 J_0\left(\ri^{3/2} \alpha \right)^2}.
\end{align}\end{subequations}
Assuming that $\beta St_f = \mathcal{O}(\beta)$ asymptotically, collecting $\mathcal{O}(1)$ problem is again Eq.~\eqref{eq:p0_pde_channel} subject to the BCs~\eqref{eq:p0_bcs_channel}, the solution of which is Eq.~\eqref{eq:P0_chan}. Next, collecting $\mathcal{O}(\beta)$ terms in Eq.~\eqref{eq:p_pde_expanded_tube} yields
\begin{equation}\label{eq:p1_pde_tube}
    \mathfrak{g}_0({\alpha})\frac{\partial^2 P_1}{\partial Z^2} = -\frac{\partial}{\partial Z}\left[\mathfrak{g}_1({\alpha})\Real[P_0]\frac{\partial P_0}{\partial Z}\right] + {2 St_f} \frac{\partial P_0}{\partial T} 
\end{equation}
subject to the same homogeneous BCs as in Eq.~\eqref{eq:p1_bcs_channel}. The solution for the first-order correction is found, from  Eqs.~\eqref{eq:p1_pde_tube} and \eqref{eq:p1_bcs_channel}, to be:
\begin{equation}
    P_1(Z,T) = \frac{1}{6} Z(1-Z)\left[3\frac{\mathfrak{g}_1({\alpha})}{\mathfrak{g}_0({\alpha})}\Real[\re^{2\pi \ri T}]\re^{2\pi \ri T} + (Z-2) \frac{2 St_f}{\mathfrak{g}_0({\alpha})} 2\pi \ri \re^{2\pi \ri T}\right].
    \label{eq:P1_tube}
\end{equation}

The perturbative, real-valued pressure distribution is then found from Eqs.~\eqref{eq:P0_chan} and \eqref{eq:P1_tube} as $\Real[P_0(Z,T)] + \beta \Real[P_1(Z,T)]$. 
Finally, using the cycle averaging defined in Eq.~\eqref{eq:cycle_avg}, we find that the real-valued cycle-averaged pressure is
\begin{equation}\label{eq:P_avg_tube}
    \begin{aligned}
    \langle P \rangle(Z) = \Real[\underbrace{\langle P_0 \rangle}_{=0} + \beta\langle P_1 \rangle] + \mathcal{O}(\beta^2)
    &= \frac{\beta}{4} Z(1-Z) \Real\left[\frac{\mathfrak{g}_1({\alpha})}{\mathfrak{g}_0({\alpha})}\right] + \mathcal{O}(\beta^2)\\
    &\simeq \frac{\beta}{4} Z(1-Z) \left(4 - \frac{17}{288}{\alpha}^4\right) + \mathcal{O}(\beta{\alpha}^8,\beta^2),
    \end{aligned}
\end{equation}
where we have given the ${\alpha}\ll1$ expansion for completeness. As before, $\Real[\mathfrak{g}_1({\alpha})/\mathfrak{g}_0({\alpha})]$ in Eq.~\eqref{eq:P_avg_tube} is evaluated numerically for plotting. Observe that Eq.~\eqref{eq:P_avg_tube} has the same form as Eq.~\eqref{eq:P_avg_channel} for a channel, save for the different dependence on $\alpha$, highlighting the ``universal'' nature of the streaming phenomenon in compliant conduits. 

Evidently, Eq.~\eqref{eq:P_avg_tube} implies the existence of a steaming pressure gradient ${\partial\langle P \rangle(Z)}/{\partial Z}$, which engenders a nonzero mean flow rate $\langle{Q}\rangle$. From the dimensionless flow rate corresponding to  Eq.~\eqref{eq:Worms_flow_rate}, a lengthy, but straightforward, calculation shows that
\begin{equation}
    \langle Q\rangle 
    = \frac{\beta}{4 } \Real\left[ \mathfrak{g}_1 (\alpha) \right] + \mathcal{O}(\beta^2)\\
    = \frac{\beta}{8} \left(1 -\frac{11}{1536}\alpha^4\right) + \mathcal{O}(\beta\alpha^6,\beta^2),
    \label{eq:Qavg_tube}
\end{equation}
which is evidently independent of $Z$. 
Furthermore, unlike the 2D channel result in Eq.~\eqref{eq:Qavg_chan}, $\langle Q\rangle/\beta$ from Eq.~\eqref{eq:Qavg_tube} is a monotonically decreasing function of $\alpha$, decaying to zero as $\alpha\to\infty$. As a consistency check, note that the leading $\beta/8$ term in Eq.~\eqref{eq:Qavg_chan} matches the $\alpha=0$ case analyzed in \cite{AC20}, specifically Eq.~(90) therein (upon neglecting wall inertia and fluid compressibility, and simplifying).

\subsection{Numerical results and discussion}\label{sec:Results_Tube}

We performed 3D direct numerical simulation using svFSI, a solver within the open-source cardiovascular modeling software \textsc{SimVascular} \cite{Upde17,Lan18}. SvFSI uses the arbitrary Eulerian--Lagrangian framework in the finite element method to solve the two-way coupled FSI problem in a monolithic approach \cite{svFSI}. The large-deformation `Saint-Venant--Kirchhoff' solid model was used in svFSI. The 3D, unsteady incompressible Navier--Stokes equations are solved in the fluid domain without the assumption of axisymmetry. Following \cite{WPC22}, the simulation was set up by creating a cylindrical fluid domain surrounded by an elastic solid mesh representing a thin tube, maintaining an aspect ratio of $\epsilon=0.0667$. For the simulations, a conforming unstructured mesh was used for each of the fluid and solid domains, with a combined $160,859$ tetrahedral elements, which was created in the commercial software ANSYS and converted to a format compatible with svFSI. An oscillatory input was given by a time-varying boundary condition at the inlet, specifically a cosine variation of the pressure, whose amplitude was matched to yield $\beta=0.05$ and $St_f=2$ and whose frequency was matched to yield the desired value of $\alpha$. (Note that, since we are considering a thin elastic tube, $\beta$ does not have to be too large to observe nonlinear effects due to FSI, and indeed it cannot be too large before nonlinear deformations of the tube itself emerge; see the discussion in \cite{AC18b}.)

Three different svFSI simulations corresponding to Womersley numbers such that ${\alpha}^2 = 1$, $3$, and $5$ were performed with a time step of $10^{-7}~\si{\second}$ for 60 periods of the forcing. We verified this time step is sufficient to resolve the oscillatory flow and fluid--structure interaction (decreasing the simulation time step does not change the results shown). The simulation data was saved in dimensionless time steps of $\Delta T = 0.1$, which is the same interval used to compute the cycle-averaged pressure. We verified that $\Delta T = 0.1$ is sufficient for the time-averaging (decreasing this $\Delta T$ does not change the averaged result). An additional simulation for ${\alpha}^2=1$ with $\beta=0.005$ was performed to compare the 3D simulation results with the perturbation solution in the weakly deformable regime.

\begin{figure}
    \centering
    \begin{subfigure}[b]{0.49\textwidth}
         \centering
         \includegraphics[width=\textwidth]{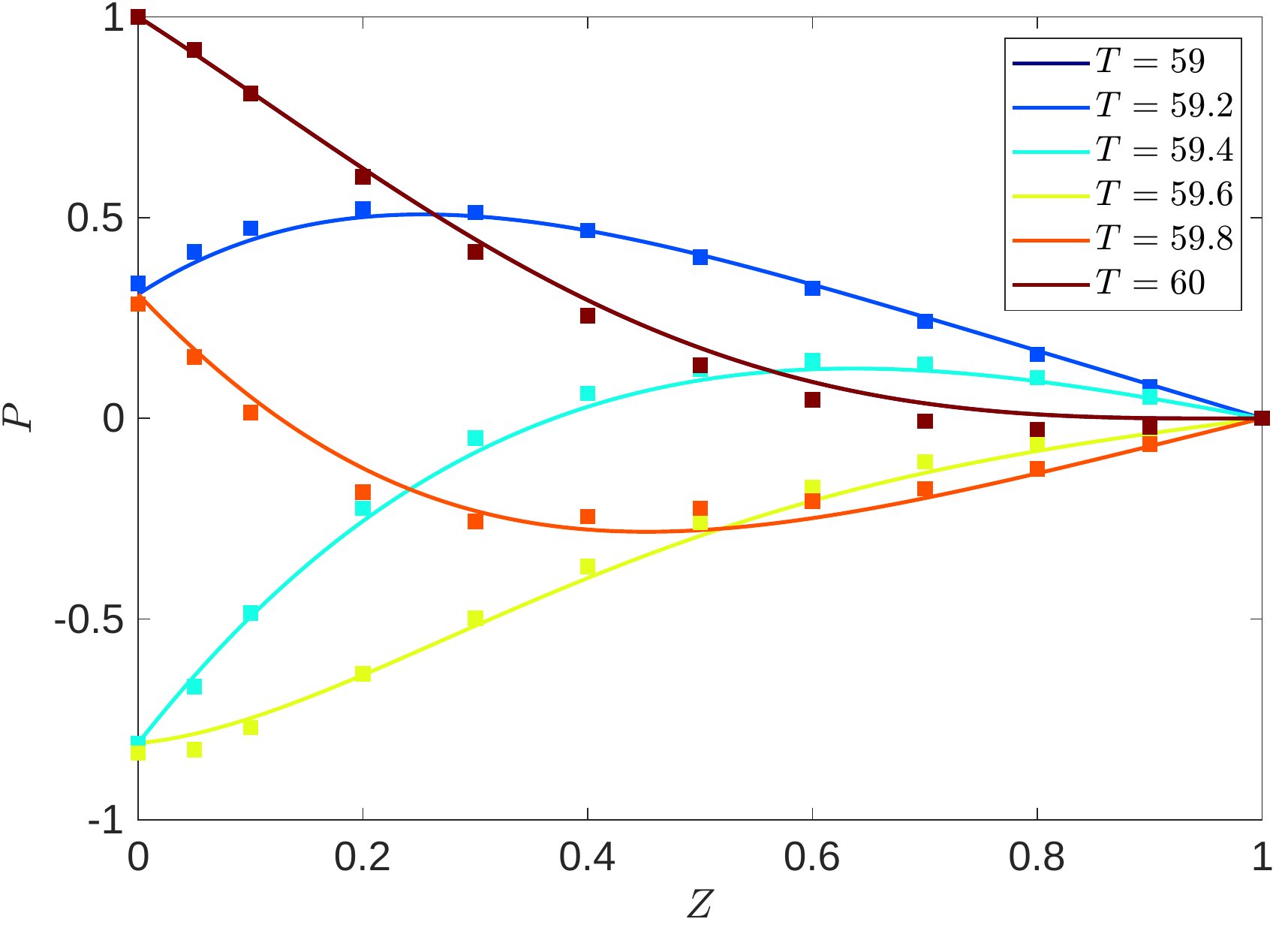}
         \caption{}
         \label{fig:pressure_tube_wo_1}
     \end{subfigure}
     \hfill
     \begin{subfigure}[b]{0.49\textwidth}
         \centering
         \includegraphics[width=\textwidth]{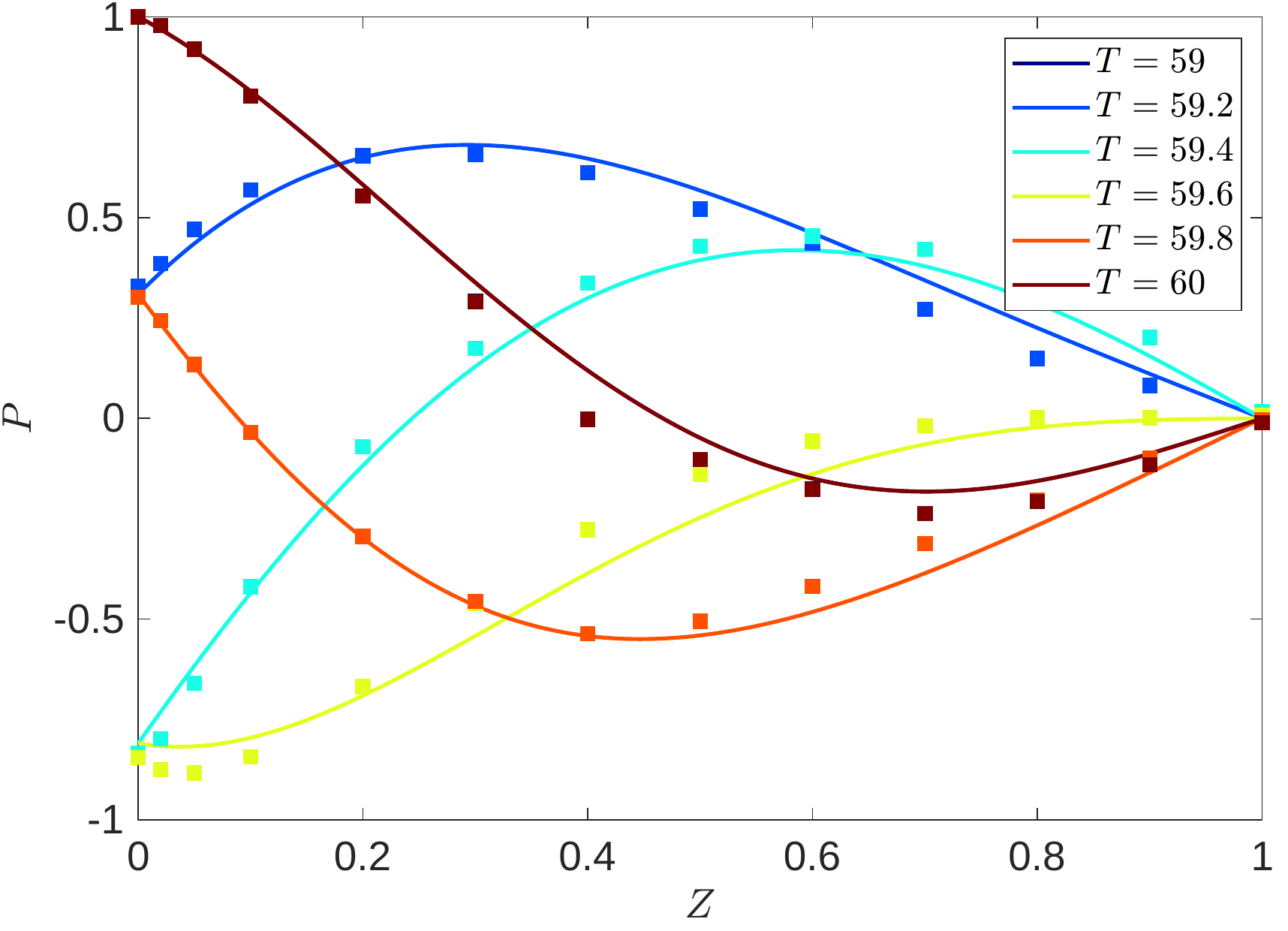}
         \caption{}
         \label{fig:pressure_tube_wo_5}
     \end{subfigure}
     \begin{subfigure}[B]{0.49\textwidth}
         \includegraphics[width=\textwidth]{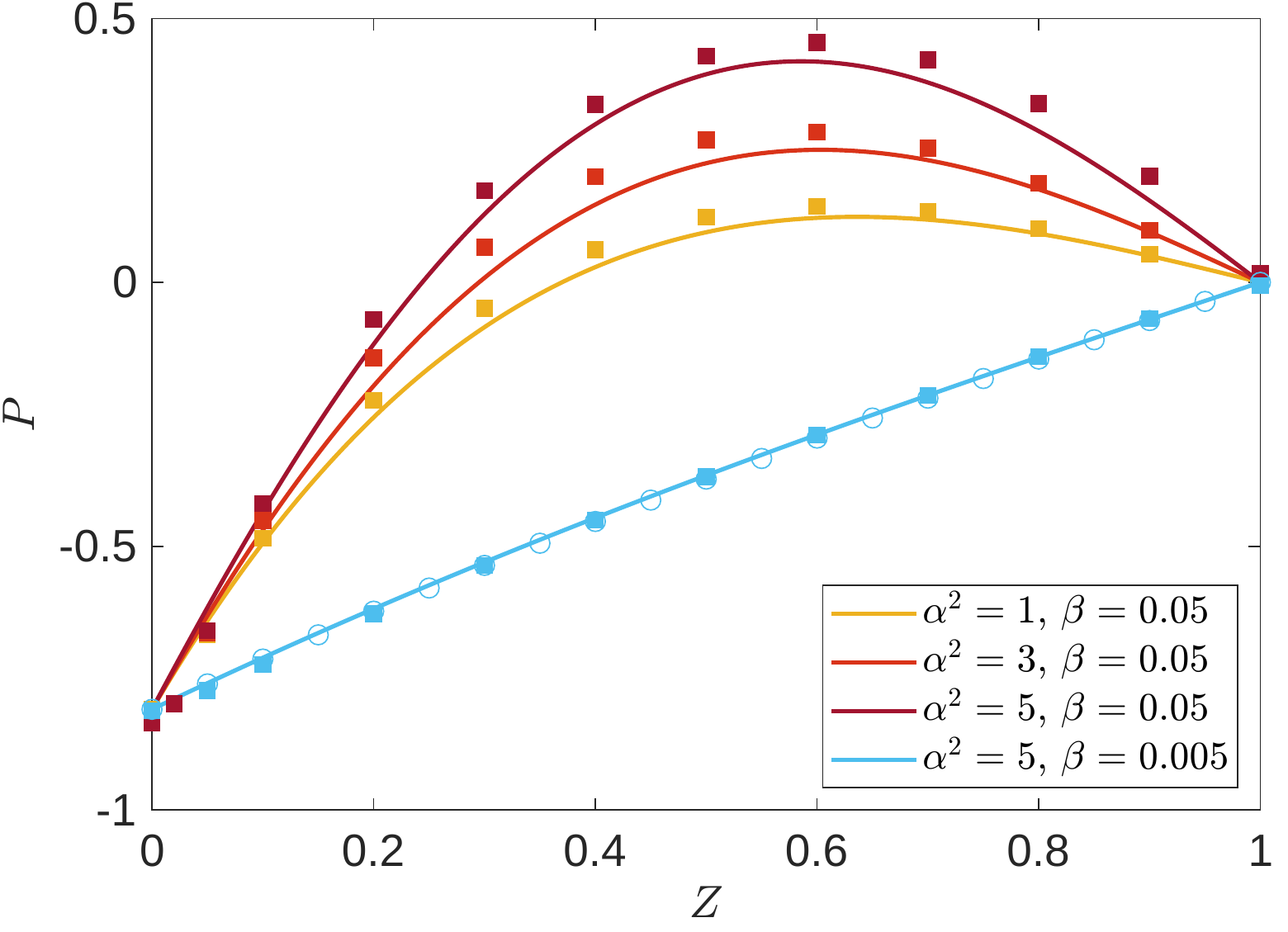}
         \caption{}
         \label{fig:pressure_dist_tube}
     \end{subfigure}
     \begin{subfigure}[b]{0.49\textwidth}
          \includegraphics[width=\textwidth]{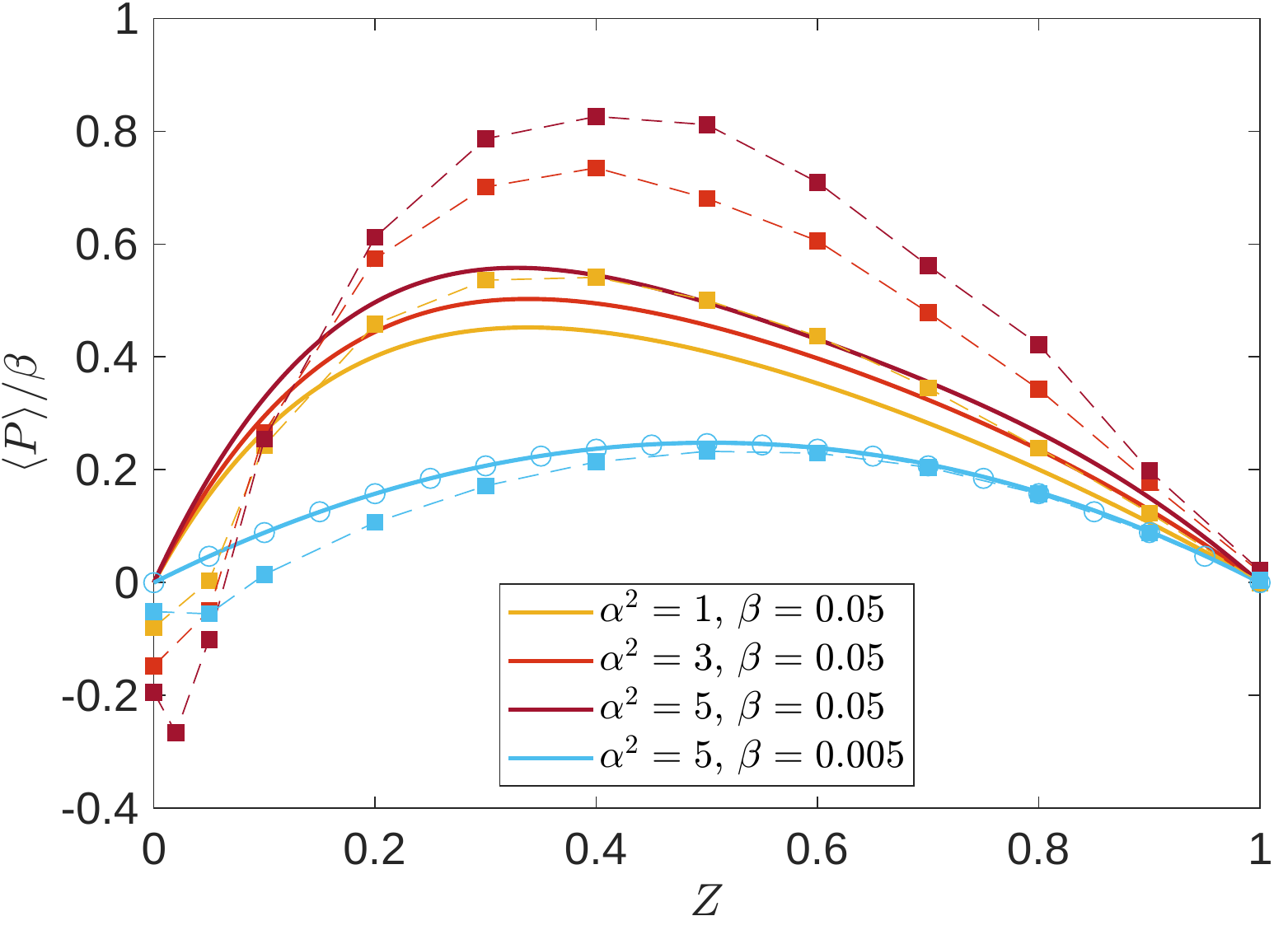}
         \caption{}
         \label{fig:streaming_tube}
      \end{subfigure}
    \caption{Dimensionless pressure distributions' evolution over a cycle for (a) $\alpha^2=1$ and (b) $\alpha^2=5$; for both, $\beta=0.05$ and $St_f=2$. (c) The dimensionless pressure distribution at the dimensionless time of $T=59.4$ for different values of the Womersley number $\alpha$ and the compliance number $\beta$. (d) The `universal' normalized streaming pressure $\langle P \rangle/\beta$ profile for different values of $\alpha$ and $\beta$. In all panels, square ($\blacksquare$) symbols denote the results from svFSI simulations, while solid curves denote the numerical solution of Eq.~\eqref{eq:dimless_P_pde} subject to Eqs.~\eqref{eq:p_ic} and \eqref{eq:p_bc}. In (c), circle ($\circ$) symbols represent the analytical perturbation solution $\Real[P_0+\beta P_1]$ found from Eqs.~\eqref{eq:P0_chan} and \eqref{eq:P1_tube}, while in (d), they represent the analytical perturbation solution found from Eq.~\eqref{eq:P_avg_tube}. Dashed lines connect symbols as a guide to the eye.}
    \label{fig:P_tube}
\end{figure}

The proposed reduced-order model~\eqref{eq:dimless_P_pde} for the pressure, which we remind is non-perturbative in $\beta$ (having taken into account two-way FSI coupling), is solved numerically in \textsc{Matlab} using \texttt{pdepe} using the same settings as in Sec.~\ref{sec:results_chan}.
Upon obtaining the numerical solution, we take its real part and compare it to the direct numerical simulation from svFSI, over an oscillation cycle, in Figs.~\ref{fig:pressure_tube_wo_1} and \subref{fig:pressure_tube_wo_5} at $\alpha^2=1$ and $\alpha^2=5$, respectively. Meanwhile, Fig.~\ref{fig:pressure_dist_tube} shows the same comparison at fixed $T=59.4$ for different Wormersley numbers. We observe that the instantaneous pressure profiles from the reduced-order model follow the 3D simulation data closely. The perturbative analytical solution from Eqs.~\eqref{eq:P0_chan} and \eqref{eq:P1_tube} is also shown for  ${\alpha}^2=1$ in Fig.~\ref{fig:pressure_dist_tube}, and it also agrees well with the 3D simulation. These comparisons demonstrate that the reduced-order model~\eqref{eq:dimless_P_pde} accurately captures the two-way-coupled fluid--structure interaction across a range of Womersley numbers.

Figure~\ref{fig:streaming_tube} shows the corresponding comparisons, for different values of the Womersley number, of the normalized steaming pressure profile $\langle P \rangle/\beta$. The streaming pressure is computed again in \textsc{Matlab} using \texttt{trapz} with a dimensionless time step of $\Delta T =0.1$ from the \texttt{pdepe} solution of Eq.~\eqref{eq:dimless_P_pde}. Meanwhile, the 3D pressure solutions from svFSI were also averaged over 10 times points of the cycle at several different cross-sections along the length of the tube and made dimensionless using the scales from Table~\ref{table:scales_AS}. Figure~\ref{fig:streaming_tube} also shows the perturbation solution from Eq.~\eqref{eq:P_avg_tube} for the smallest value of the compliance number $\beta$. We observe qualitative agreement between the streaming pressure profiles. The quantitative agreement is not as good as for the pressure profiles themselves. This discrepancy can be attributed to the fact that streaming is a weak effect and $\langle P \rangle$ is on the order of the small differences between the 3D simulations and the reduced model already present in Figs.~\ref{fig:pressure_tube_wo_1} and \subref{fig:pressure_tube_wo_5}. Therefore, these small discrepancies are exaggerated in Fig.~\ref{fig:streaming_tube}. Weak elasticity effects, neglected in our reduced-order model (recall Sec.~\ref{sec:deformation_model_tube}), are captured by the svFSI simulations and could also be exaggerated when considering $\langle P\rangle/\beta$. Further, $\langle P \rangle$ near $Z=0$ is not exactly zero, as would be expected from the boundary condition applied, due to the ``weak'' enforcement of the pressure boundary condition in the finite-element method \cite{BTT13}. Nevertheless, the trends and shapes of the streaming pressures, with respect to the Womersley number, from both 3D simulations and the reduced-order model agree. And, for the smallest values of $\beta$ (weakest FSI), we obtain the best agreement between the two. Although there appears to be an inlet effect in the 3D simulation data shown in Fig.~\ref{fig:streaming_tube}, given the small values of streaming pressure we are dealing with, we cannot attach any physical significance to this observation.

\section{Conclusion}
\label{sec:conclusion}

For the two canonical geometries of a 2D channel and a 3D axisymmetric tube, we derived reduced-order, one-dimensional (1D) models, Eqs.~\eqref{eq:p_pde_channel} and \eqref{eq:dimless_P_pde} respectively, for the unsteady pressure variation due to oscillatory flow in compliant conduits. The key consideration in our models, improving upon previous work on fluid--structure interactions due to oscillatory internal flows, is to incorporate two-way coupling. Specifically, the flow-induced deformation of the compliant conduit affects the pressure distribution and \textit{vice versa}. 
Since two-way coupling forestalls an analytical solution of the fluid's momentum equation, we motivated a von K\'{a}rm\'{a}n--Pohlhausen-type closure using the known axial velocity profiles (and volumetric flow rate) for pulsatile flow using Womersley's eponymous solution. Importantly, in our model, the pressure (and, thus, pressure gradient) cannot be specified \textit{a priori} as is done in one-way coupled fluid--structure interactions.

Our model makes no \textit{a priori} assumptions on the relative importance of unsteady inertial forces over viscous forces in the flow (quantified by the Womersley number) and was indeed demonstrated to be valid over a wide range of Womersley numbers. Although the model is a nonlinear PDE with no analytical (or even separable) solutions, we made some analytical progress in the limit of weakly deformable conduits. In doing so, we demonstrated that the cycle-averaged pressure, normalized by a suitable fluid--structure interaction parameter, is both non-vanishing and a universal function of the spatial coordinate and Womersley number. This analytical result, confirmed by both simulations of the governing 1D PDE and 3D direct numerical simulations (in the case of an axisymmetric tube), shows that viscous streaming can be self-induced by fluid--structure interactions in internal incompressible flows. In the more realistic case of a 3D axisymmetric tube, we validated the 1D reduced model against direct numerical simulations performed using the open-source software \textsc{SimVascular}, specifically its svFSI solver. The 3D simulation results for the pressure variation agree well with the 1D reduced model, as well as its analytical solutions for weakly deformable conduits, demonstrating the validity of our theory of oscillatory flows in deformable conduits at arbitrary Womersley number.

Although we focused on zero-mean oscillatory flows, our models can also be used to study pulsatile flows in which there is a nonzero mean pressure (flow) component on top of the zero-mean oscillating pressure, by simply changing the boundary condition at the inlet. Such pulsatile flows find applications to inertial focusing of particles in microfluidics \cite{MET18,VJ20,VJ21}, and are of interest to understanding, \textit{e.g.}, the biophysics of endothelial cells in the cardiovascular system, which experience flows driven by the pulsations of the heart \cite{DDS20}. Although compliance of the conduit has not received much attention in the context of particle migration in microfluidics (notably, it is absent in the recent review \cite{SD19}), wall compliance has been shown to provide a way to sieve particles \cite{CFZZWD20} and blood cells \cite{CZGLWD22} in microchannels. Thus, our model paves the way to size-selective particle migration/sorting via pulsatile flows in compliant conduits.

It would also be worth exploring if our reduced models can be used to obtain a deeper understanding of the hydraulic compliance and impedance of deformable conduits, given that the current expressions for these lumped model parameters are not always accurate (as discussed in Sec.~\ref{sec:intro}). It may be that simple models based on ODEs arise as distinguished limits of a more general theory that takes into account two-way-coupled fluid--structure interactions (and thus the \emph{pressure-dependent} nature of the hydraulic resistance, capacitance, and inductance). 

On the mathematical side, in future work, it would be of interest to revisit the small-$\beta$ results above using a dual application of the reciprocal theorems for Stokes flow and linear elasticity \cite{BSC22}, providing an independent way to calculate the streaming effect.

\section*{Acknowledgements}
We would like to thank T.~C.\ Shidhore for assistance with the setup of svFSI simulations. The latter simulations were performed using the community clusters of the Rosen Center for Advanced Computing at Purdue University.
We would also like to thank E.\ Boyko for extensive discussions on oscillatory flows and lubrication theory.
I.C.C.\ would like to acknowledge the hospitality of the University of Nicosia, Cyprus, where this manuscript was completed thanks to a Fulbright U.S.\ Scholar award from the U.S.\ Department of State.
S.D.P.\ and I.C.C.\ further acknowledge partial support from the U.S.\ National Science Foundation under Grant No.\ CMMI-2029540 during the completion of this work.

SimVascular case files for the 3D simulations and MATLAB scripts implementing the reduced-order model and generating the plots in this paper are available on the Purdue University Research Repository at \url{https://dx.doi.org/10.4231/SW73-D037}.

\appendix*
\section{Limit of steady flow}

\subsection{2D Channel}
Taking the limit ${\alpha} \rightarrow 0$ of Eq.~\eqref{eq:p_pde_channel}, we obtain an ordinary differential equation for the steady pressure distribution in a 2D channel with a deformable wall:
\begin{equation}
    \frac{\rd }{\rd Z} \left\{ - \frac{\rd P }{\rd Z}\left[\frac{1}{12}(1+ \beta P)^3\right]\right\} = 0.
\end{equation}
With the boundary conditions $P(0)=1$ and $P(1)=0$, we obtain the solution for the pressure as:
\begin{equation}
    P(Z) = \frac{1}{\beta} \left(\left\{(1+\beta)^4-[(1+\beta)^4-1]Z\right\}^{1/4} - 1\right).
    \label{eq:P_steady_channel}
\end{equation}
As expected \cite{RK72}, the pressure distribution~\eqref{eq:P_steady_channel} is not linear in the deformable channel at steady state ($\rd P/\rd Z \ne const.$), but $\lim_{\beta\to0}P(Z) = 1-Z$ as usual.

Note that we cannot compare the $\alpha\to0$ limit of  Eq.~\eqref{eq:P_avg_channel} to the $\beta\ll1$ expansion of Eq.~\eqref{eq:P_steady_channel} (\textit{i.e.}, the $\alpha\to0$ and $\beta\to0$ limits do not commute) because in Eq.~\eqref{eq:p0_z0_bc_channel} we imposed a time-dependent boundary condition on $P_0$, while $P$ in Eq.~\eqref{eq:P_steady_channel} satisfies a time-independent one.

\subsection{Axisymmetric tube}
Taking the limit ${\alpha} \rightarrow 0$ of Eq.~\eqref{eq:dimless_P_pde}, we obtain an ordinary differential equation for the steady pressure distribution in an axisymmetric deformable tube:
\begin{equation}
    \frac{\rd }{\rd Z} \left\{ - \frac{\rd P }{\rd Z}\left[\frac{1}{8}(1+ \beta P)^4\right] \right\} = 0.
\end{equation}
With the boundary conditions $P(0)=1$ and $P(1)=0$, we obtain the solution for the pressure as:
\begin{equation}
    P(Z) = \frac{1}{\beta} \left(\left\{(1+\beta)^5-[(1+\beta)^5-1]Z\right\}^{1/5} - 1\right).
    \label{eq:P_steady_tube}
\end{equation}
Observe that the latter is simply the solution for the pressure in a deformable tube in the pressure-controlled regime, complementing the solution for the flow-controlled presented in \cite{AC18b}. As expected \cite{RK72}, the pressure distribution~\eqref{eq:P_steady_tube} is not linear in the deformable tube at steady state ($\rd P/\rd Z \ne const.$), but $\lim_{\beta\to0}P(Z) = 1-Z$ as usual.

Again, we cannot compare the $\alpha\to0$ limit of Eq.~\eqref{eq:P_avg_tube} to the $\beta\ll1$ expansion of Eq.~\eqref{eq:P_avg_tube} because the $\alpha\to0$ and $\beta\to0$ limits do not commute. 

\bibliography{references.bib}

\end{document}